\documentclass[english]{IEEEtran}
\usepackage[T1]{fontenc}
\usepackage[latin9]{inputenc}
\usepackage{float}
\usepackage{units}
\usepackage{amsmath}
\usepackage{amssymb}
\usepackage{graphicx}
\usepackage{esint}

\makeatletter

\floatstyle{ruled}
\newfloat{algorithm}{tbp}{loa}
\providecommand{\algorithmname}{Algorithm}
\floatname{algorithm}{\protect\algorithmname}

\makeatother

\usepackage{babel}
\begin{document}

\title{Sequential and Incremental Precoder Design for Joint Transmission
Network MIMO Systems with Imperfect Backhaul%
\thanks{Copyright (c) 2012 IEEE. Personal use of this material is permitted.
However, permission to use this material for any other purposes must
be obtained from the IEEE by sending a request to pubs-permissions@ieee.org.%
}
}

\author{{\normalsize Ming Ding, }\emph{\normalsize Member, IEEE}{\normalsize ,
Jun Zou, Zeng Yang, Hanwen Luo, and Wen Chen, }\emph{\normalsize Senior
Member, IEEE}%
\thanks{Ming Ding, Jun Zou, Hanwen Luo and Wen Chen are with the Dept. of
Electronic Engineering, Shanghai Jiaotong University, Shanghai, P.
R. China. (E-mail: \{dm2007, zoujun, hwluo, wenchen\}@sjtu.edu.cn). %
}
\thanks{Ming Ding and Zeng Yang are with Sharp Laboratories of China Co.,
Ltd. (E-mail: \{ming.ding, zeng.yang\}@cn.sharp-world.com).%
}
\thanks{This work is sponsored by Sharp Laboratories of China Co., Ltd.%
}}

\maketitle
{}
\begin{abstract}
In this paper, we propose a sequential and incremental precoder design
for downlink joint transmission (JT) network MIMO systems with imperfect
backhaul links. The objective of our design is to minimize the maximum
of the sub-stream mean square errors (MSE), which dominates the average
bit error rate (BER) performance of the system. In the proposed scheme,
we first optimize the precoder at the serving base station (BS), and
then sequentially optimize the precoders of non-serving BSs in the
JT set according to the descending order of their probabilities of
participating in JT. The BS-wise sequential optimization process can
improve the system performance when some BSs have to temporarily quit
the JT operations because of poor instant backhaul conditions. Besides,
the precoder of an additional BS is derived in an incremental way,
i.e., the sequentially optimized precoders of previous BSs are fixed,
thus the additional precoder plays an incremental part in the multi-BS
JT operations. An iterative algorithm is designed to jointly optimize
the sub-stream precoder and sub-stream power allocation for each additional
BS in the proposed sequential and incremental optimization scheme.
Simulations show that, under the practical backhaul link conditions,
our scheme significantly outperforms the autonomous global precoding
(AGP) scheme in terms of BER performance.\end{abstract}

\section{Introduction}

Recently, there have been considerable interests in network multiple-input
multiple-output (MIMO) systems \cite{CoMP_overview[1]}, where multiple
geographically distributed multi-antenna base stations (BSs) cooperate
with each other to transmit data to users. In \cite{ZF-JT-N-MIMO[2]},
it is shown that network MIMO can be employed to mitigate co-channel
interference and to exploit macro-diversity. Motivated by these works,
downlink network MIMO technologies have been adopted by 4G mobile
communication standards, such as the 3rd Generation Partnership Project
(3GPP) Long Term Evolution-Advanced (LTE-A) networks \cite{LTE-A[3]}.

The cooperating strategies for network MIMO systems can be generally
divided into two categories, i.e., coordinated beamforming (CB) and
joint transmission (JT). When the CB strategy is employed, the cooperating
BSs share channel state information (CSI) in various forms, and each
BS only transmits data to its own served users, while for the JT strategy,
both CSI and user data should be shared by all cooperating BSs, and
thus each user receives data from multiple BSs. Many current research
works focus on CB network MIMO schemes \cite{CB1[4],CB2[5],CB3[6],CB4[7]}.
However, the performance of CB strategy is generally interference-limited
due to lack of abundant spatial-domain degrees of freedom for perfect
inter-BS interference coordination in practical systems \cite{CoMP_overview[1]}.
On the other hand, the JT strategy takes a more aggressive approach
to cope with the interference problem by transforming the interference
from neighbor BSs into useful signals. Usually, the mathematical form
of the JT scheme bears a close resemblance to conventional MIMO systems
except for its distributed structure \cite{JT-BC[8]}. From uplink-downlink
duality theory, the capacity region of the downlink JT network MIMO
systems can be computed from its dual uplink \cite{Duality[9]} with
the same sum power constraint. These results were later generalized
to accommodate the per-antenna power constraint \cite{YuWei_perAnte[10]}
by showing that the per-antenna downlink transmitter optimization
problem can be transformed into a dual uplink problem with an uncertain
noise. It should be noted that most capacity duality results are based
on non-linear signal processing at the BS side, such as the dirty
paper coding \cite{DPC[11]}, which is computationally demanding for
precoding across multiple BSs. This has motivated research in linear
precoding for JT, such as zero-forcing (ZF) precoders \cite{ZF-JT[12]},
which are much more easy-to-implement compared with the DPC precoder.
Moreover, when user equipment (UE) is equipped with multiple antennas,
the distributed transceiver design \cite{transceiver_design[13]},
i.e., designing the precoder with UE\textquoteright{}s receiver structure
taken into account, should also be considered for JT. Another relevant
issue regarding JT is the imperfect backhaul links \cite{imperfect_backhaul[14]}.
In practice, cooperating BSs are connected through imperfect links
with finite capacity, unpredictable latency, and limited connectivity.
For example, the latency of practical backhaul links such as the copper
and wireless interface varies from several milliseconds to tens of
milliseconds depending on technology/standard. Besides, when the backhaul
communication is based on a generic IP network, the backhaul latency
also depends on the number of routers between two cooperative BSs
and the topology of the network, e.g., star, ring, tree, mesh, etc.
Furthermore, congestion in the routers causes an extra delay typically
of several milliseconds \cite{backhaul_orange[15]}. It should be
noted that limited capacity is another important backhaul issue \cite{Backhaul Capacity[16]}.
Most of the current cellular backhaul networks are designed for handover
functions, which are not suited for data exchange in large amount
\cite{LTE-A[3]}. Constraints from lower capacity/higher latency backhaul
communication in coordinated multi-point (CoMP) operations were studied
in the 3GPP LTE-A meetings \cite{CoMP_backhaul_discus_63bis[17]},
\cite{CoMP_backhaul_discus_64[18]}. Due to the immature status of
the study, remote radio head (RRH) or remote radio equipment (RRE)
based centralized BS and fiber based backhaul \cite{fiber_backhaul[19]}
were assumed for the JT as a starting point of the working order for
the imperfect backhaul issue \cite{CoMP_rev_SID[20]}. Although the
current centralized network structure and fiber based backhaul will
not pose serious problems for existing JT schemes, for future JT operations,
the impact of imperfect backhaul should be carefully investigated
\cite{investigate_future_backhaul[21]}. Up to now, theoretical performance
bounds for JT network MIMO with unreliable backhaul links among the
cooperating BSs are unknown yet \cite{CoMP_overview[1]}. Finally,
it should also be noted that imperfect CSI fed back by the UE is a
common assumption in practical frequency division duplexing (FDD)
systems such as the 3GPP LTE-A system, where the downlink CSI cannot
be inferred from the uplink CSI. For imperfect CSI feedback in a practical
system, implicit CSI feedback, i.e., feedback of precoder recommendation
by UEs \cite{LTE-A[3]}, is much more preferred than the explicit
CSI feedback, i.e., feedback of the channel matrix, due to feedback
overhead considerations. In this paper, we consider the implicit CSI
feedback, which has been widely adopted by practical systems such
as the 3GPP LTE-A system \cite{LTE-A[3]}.

In this paper we investigate the precoder design for downlink JT network
MIMO systems with imperfect backhaul links, i.e., finite capacity,
unpredictable latency, and limited connectivity. In particular, we
focus on network impairments incurred by backhaul delays. For JT operations,
both the transmission data and the CSI must be available at the actual
transmission BSs before the transmission starts; otherwise the JT
cannot be operated as it was supposed to be. In a multi-BS JT network
where one serving BS and several helper BSs constitute a JT set, we
assume that UE\textquoteright{}s data will always arrive at the serving
BS from higher-layer entities in a timely and error-free manner, then
the serving BS shares the data with the helper BSs by means of imperfect
backhaul communications. Here, the serving BS does not necessarily
mean the BS with the strongest signal level at the UE, the typical
event of which occurs during inter-BS handover process. According
to \cite{LTE-A[3]}, the serving BS is defined as the BS sending downlink
control signalings, e.g., downlink scheduling information, to the
UE. In the sense of downlink control signaling connection, the helper
BSs are not equal partners with the serving BS because they only provide
data transmissions to the UE. Besides, regarding the CSI feedback,
we assume that the UE reports the precoder recommendation information
either to each individual BS in the JT set or to the serving BS which
in turn exchanges this information among helper BSs over backhaul
links. Whether both the transmission data and the CSI arrive at a
certain helper BS before the JT scheduled to be performed or not is
a probabilistic event because of the non-deterministic delay, which
has been shown to conform to a shifted gamma distribution in \cite{backhaul_sgamma[22]}.
If a helper BS fails to obtain both the transmission data and the
CSI in time, then it has to quit the JT operations. Thereby, whether
a helper BS can participate in JT or not is also a probabilistic event.
Since the average bit error rate (BER) of the system is generally
dominated by the sub-stream with the maximum mean square error (MSE)
\cite{motivation_MINMAX[15]}, we propose a precoding scheme to minimize
the maximum of the sub-stream MSEs. In the proposed scheme, we first
optimize the precoder at the serving BS, and then sequentially optimize
the precoders of helper BSs in the JT set according to the descending
order of their probabilities of participating in JT. The BS-wise sequential
optimization process can improve the system performance when some
helper BSs have to temporarily quit the JT operations because of poor
instant backhaul conditions. Besides, the precoder of an additional
BS is derived in an incremental way, i.e., the sequentially optimized
precoders of previous BSs are fixed, thus the additional precoder
plays an incremental part in the multi-BS JT operations. Because the
BS precoders are generated sequentially and incrementally, our proposed
scheme will be referred to as the sequential and incremental precoding
(SIP) scheme hereafter. An iterative algorithm is designed to jointly
optimize the sub-stream precoder and power allocation (PA) for each
additional BS in the SIP scheme. Simulation results show that our
scheme can achieve considerable gains in terms of BER performance
compared with a variation of global precoding (GP) scheme, i.e., autonomous
global precoding (AGP) scheme, under the practical backhaul link conditions.

There are several benefits offered by our scheme. First, it offers
flexibility to the JT network MIMO since the helper BSs can adaptively
decide whether or not to join JT according to their own situation.
Such JT scheme enables the network MIMO to adaptively switch among
single-BS transmission (ST), partial JT and full JT without inter-BS
signaling. Here partial JT refers to the transmission from a subset
of BSs within the JT set. Second, the related CSI feedback scheme
can easily fit into the current 3GPP LTE-A per-BS feedback framework
\cite{LTE-A[3]}, i.e., the feedback operation is performed on a per-BS
basis, which facilitates the feedback design. Third, the complexity
on the UE side to select a preferred precoder from a codebook is low.
Instead of searching for the precoder through a large codebook as
done in the conventional GP scheme, the precoder for each BS is obtained
from a small per-BS based codebook. .

The rest of the paper is organized as follows. Section II presents
the system model and briefly describes backhaul impairments. Section
III discusses the GP and Autonomous GP (AGP) scheme. Section IV proposes
the sequential and incremental precoder design and its extension to
multi-BS JT network MIMO systems. The paper is completed with simulation
results and conclusions in sections V and VI, respectively.

\textit{Notations}: $\left(\cdot\right)^{\textrm{T}}$, $\left(\cdot\right)^{\textrm{*}}$,
$\left(\cdot\right)^{\textrm{H}}$ and $\textrm{tr}\left\{ \mathbf{\cdot}\right\} $
stand for the transpose, conjugate, conjugate transpose, and trace
of a matrix, respectively. $\mathbf{A}_{i,j}$, $\mathbf{A}_{i,:}$,
and $\mathbf{A}_{:,j}$ denote the ($i$, $j$)-th entry, $i$-th
row, and $j$-th column of matrix $\mathbf{A}$, respectively. $\textrm{diag}\left\{ \mathbf{A}_{k}\right\} $
denotes a block-diagonal matrix with the $k$-th diagonal block given
by $\mathbf{A}_{k}$. $\mathbf{I}_{N}$ stands for an $N\times N$
identity matrix. $\left|\mathbf{a}\right|$ denotes the Euclidean
norm of a vector $\mathbf{a}$. $\mathbb{E}\left\{ \cdot\right\} $
and $\textrm{Re}\left\{ \cdot\right\} $ denote expectation operator
and the real part of a complex value, respectively. Finally, we define
$\left(a\right)^{+}=\max\left(0,a\right)$.

\section{System Model }

In this section, we address the system model and briefly discuss backhaul
impairments. We consider a multi-cell wireless network consisting
of $B$ adjacent BSs, where each BS is equipped with $N_{\textrm{T}}$
antennas. A cell edge UE with $N_{\textrm{R}}$ antennas is served
by one serving BS, and the other $B-1$ BSs are helper BSs which adaptively
provide service to the UE depending on the backhaul conditions. In
practice, the candidates of helper BSs can be decided either by UE
based on received reference signal strength of nearby BSs, or by the
serving BS based on wideband CSI reported by UE. The selection algorithm
to decide the helper BSs can be found in \cite{BS_selection_1[24]}
and \cite{BS_selection_2[25]}. Here, we assume $B-1$ helper BSs
have already been selected based on some existing BS selection schemes.
Moreover, in practical scenarios, values of $B$ are relatively small,
usually not larger than 4 \cite{CoMP_cluster_size[26]}.

\subsection{The JT Networks with Two BSs}

Our basic idea to optimize the precoders for the $B$ BSs in the JT
set to derive per-BS precoders one by one in a sequential and incremental
manner. Hence, the most basic scenario is a JT network with only two
BSs. A two-BS JT network $\left(B=2\right)$ is shown in Fig. 1, which
serves as an instructive example to formulate the key problem of our
concern. An $N_{\textrm{R}}$-antenna cell edge UE is associated with
a serving $\mathrm{BS}_{1}$ and the transmission is assisted by a
helper $\mathrm{BS}_{2}$. Note that the results from this model will
later be extended to a more general model with multiple BSs in the
JT set.

\begin{figure}[H]
\centering \includegraphics[width=8cm]{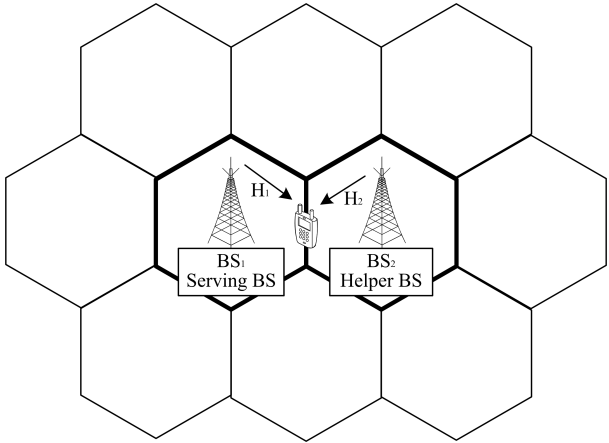} 
\\
\vspace{-0.5em}
\renewcommand{\figurename}{Fig.}

\caption{Illustration of a two-BS JT network.}
\end{figure}

In Fig. 1, the base-band channel matrix between the $b$-th BS and
the UE is denoted as $\mathbf{H}_{b}\in\mathbb{C}^{\mathit{N_{\textrm{R}}\times N_{\textrm{T}}}}$.
The UE reports the precoder recommendation information either to each
individual BS in the JT set or to the serving BS which in turn exchanges
this information among helper BSs over backhaul links. Let $\mathbf{W}_{b}\in\mathbb{C}^{\mathit{N_{\textrm{T}}\times L}}$
be the local precoding matrix of the $b$-th BS, where $L$ is the
number of independent data sub-streams for the UE. In addition, $\mathbf{W}_{b}$
is subjected to a per-BS power constraint $\textrm{tr}\left\{ \mathbf{W}_{b}^{\textrm{H}}\mathbf{W}_{b}\right\} \leq P$,
where $P$ is the maximum transmission power at each BS. Then the
signal received at the UE can be described by

\begin{equation}
\mathbf{y}=\mathbf{\left[\mathbf{H}_{\mathrm{1}},\mathbf{H}_{\mathrm{2}}\right]\left[\begin{array}{c}
\mathbf{W}_{\mathrm{1}}\\
\mathbf{W}_{\mathrm{2}}
\end{array}\right]x}+\mathbf{n},
\end{equation}
where $\mathbf{x}=\left[x_{1},x_{2},\cdots,x_{L}\right]^{\textrm{T}}$
is the transmission data vector with $\mathbb{E}\left\{ \mathbf{x\mathbf{x^{\textrm{H}}}}\right\} =\mathbf{I}_{L}$
and $\mathbf{n}$ is the noise vector with $\mathbb{E}\left\{ \mathbf{n\mathbf{n^{\textrm{H}}}}\right\} =\mathbf{R}_{\textrm{n}}$.
Note that the interference from BSs outside of the interested JT set
is incorporated into $\mathbf{n}$. Assume that the interference is
white-colored \cite{whiten_interf[27]}, then $\mathbf{R}_{\textrm{n}}$
can be simplified to $\mathbf{R}_{\textrm{n}}=N_{0}\mathbf{I}_{N_{\textrm{R}}}$.

Suppose that a linear receiver $\mathbf{F}\in\mathbb{C}^{\mathit{L\times N_{\textrm{R}}}}$
is employed at the UE to detect $\mathbf{x}$. Then the MSE of the
$i$-th $\left(i\in\left\{ 1,2,\cdots,L\right\} \right)$ detected
sub-stream can be represented by

\begin{equation}
M_{i}=\mathbb{E}\left\{ \left|\mathbf{F}_{i,:}\mathbf{y}-x_{i}\right|^{2}\right\} .
\end{equation}

In general, the average BER of the system is dominated by the sub-stream
with the maximum MSE \cite{motivation_MINMAX[15]}. Therefore, we
want to jointly design $\mathbf{W}_{1}$, $\mathbf{W}_{2}$ and $\mathbf{F}$
to minimize the maximum of the sub-stream MSEs. This MIN-MAX-MSE problem
is formulated as

\begin{equation}
\begin{array}{c}
\underset{\mathbf{F},\mathbf{W}_{1},\mathbf{W}_{2}}{\min}\quad\max\left\{ \left.M_{i}\right|i\in\left\{ 1,2,\cdots,L\right\} \right\} ,\\
\quad\textrm{s.t.}\quad\:\quad\textrm{tr}\left\{ \mathbf{W}_{b}^{\textrm{H}}\mathbf{W}_{b}\right\} \leq P,\,\forall b=1,2.
\end{array}
\end{equation}

\subsection{Modeling of Imperfect Backhaul}

For full JT operations, the transmission data vector $\mathbf{x}$
needs to be available at the actual transmission points before the
transmission starts. We assume that UE\textquoteright{}s data will
always arrive at the serving BS from higher-layer entities in a timely
and error-free manner. Then the serving BS shares the data with the
helper BSs by means of imperfect backhaul communications.

In this paper, we are mainly concerned with the impact of BS backhaul
latency on system performance. In practice, backhaul links can be
generally classified into three categories according to the physical
media, i.e. optical fiber, copper (ADSL, ATM, VDSL, etc.) and wireless
interface. The typical latency of optical fiber is below 1 ms, which
can be neglected since the usual delay of channel state information
(CSI) exchange and scheduling in cooperative MIMO systems is approximately
10 ms \cite{backhaul_orange[15]}. However, the latency of copper
and wireless interface backhaul links varies from several milliseconds
to tens milliseconds depending on technology/standard. Besides, when
the backhaul communication is based on a generic IP network, the backhaul
latency also depends on the number of routers between two cooperative
BSs and the topology of the network, e.g., star, ring, tree, mesh,
etc. Furthermore, congestion in the routers causes an extra delay
typically of several milliseconds \cite{backhaul_orange[15]}.

\begin{figure}[H]
\centering \includegraphics[width=8cm]{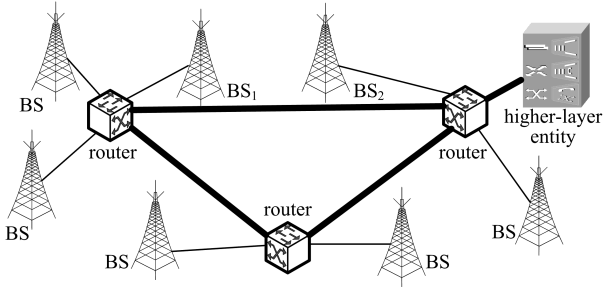} 
\\
\vspace{-0.5em}
\renewcommand{\figurename}{Fig.}

\caption{Illustration of a realistic BS backhaul network.}
\end{figure}

Fig. 2 illustrates an example of a practical BS backhaul network,
which combines the ring and tree topologies. If $\mathrm{BS}_{1}$
wants to share data with $\mathrm{BS}_{2}$, it has to set up a backhaul
link over several routers. It should be noted that limited capacity
is another important backhaul issue \cite{Backhaul Capacity[16]}.
Most of the current cellular backhaul networks are designed for handover
functions, which are not suited for data exchange in large amount
\cite{LTE-A[3]}. However, we assume capacity is adequate for backhaul
communication throughout this paper.

According to \cite{backhaul_20ms[18]}, the maximum delay for normal
backhaul links is around 20 ms and the typical average delay is expected
to be within 10 ms. More detailed description of the backhaul latency
model can be found in \cite{backhaul_sgamma[22]}. The backhaul delay
conforms to a shifted gamma distribution \cite{backhaul_sgamma[22]},
and its probability density function (PDF) can be represented by

\begin{equation}
f\left(t\right)=\frac{\left(\frac{t-t_{0}}{\alpha}\right)^{\beta-1}\exp\left\{ \frac{-\left(t-t_{0}\right)}{\alpha}\right\} }{\alpha\Gamma\left(\beta\right)},
\end{equation}
where $\alpha$, $\beta$ and $t_{0}$ are the scale, shape and shift
parameter, respectively, and $\Gamma\left(\cdot\right)$ denotes the
gamma function. According to \cite{backhaul_sgamma[22]}, the typical
values are $\alpha=1$, $\beta=2.5$ and $t_{0}=7.5$ ms. The corresponding
PDF curve is plotted in Fig. 3 for illustration purpose.

\begin{figure}[H]
\centering \includegraphics[width=8cm]{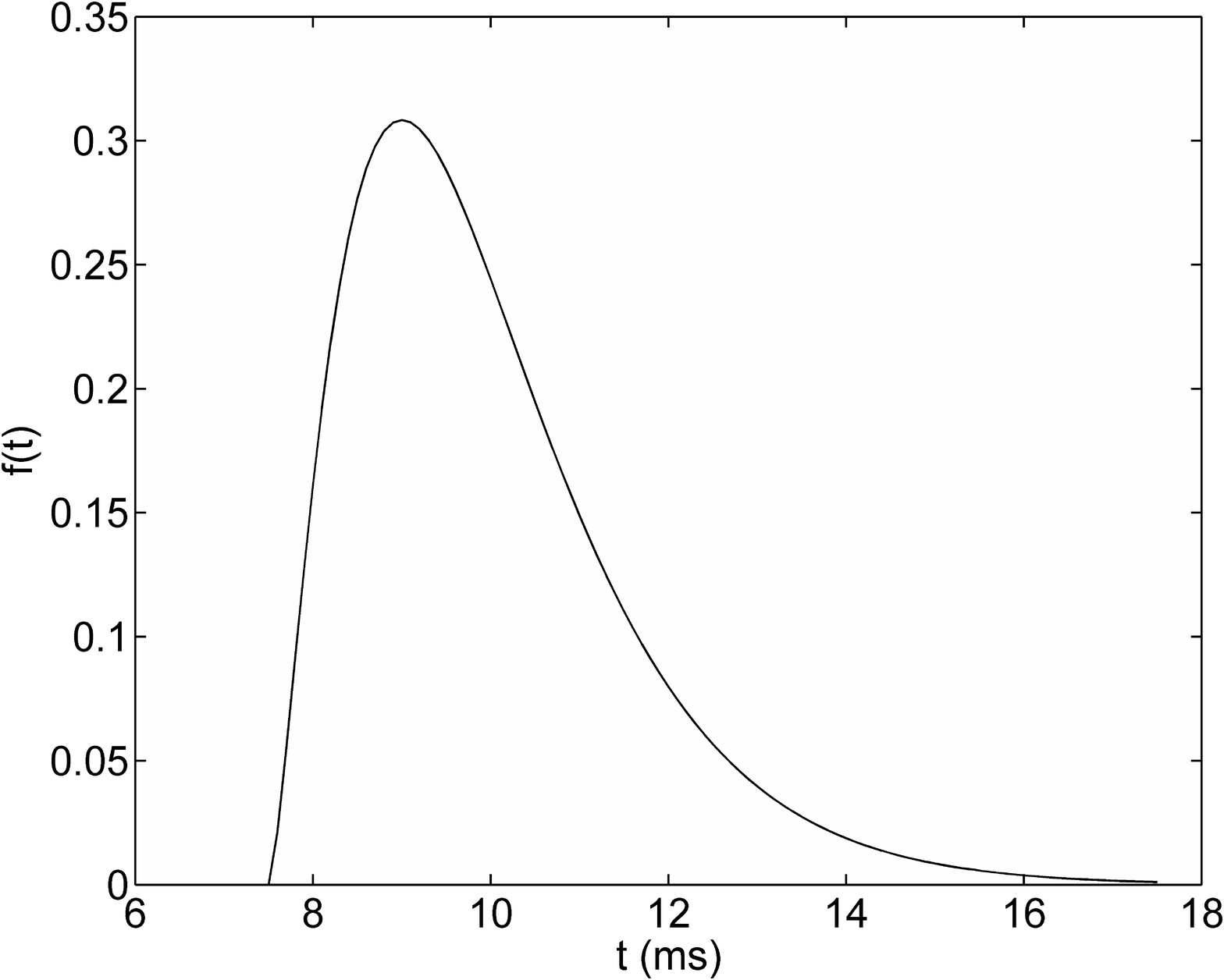} 
\\
\vspace{-0.5em}
\renewcommand{\figurename}{Fig.}

\caption{Shifted gamma distribution of the backhaul delay.}
\end{figure}

In every transmission slot, the $b$-th $\left(b\in\left\{ 2,3,\cdots,B\right\} \right)$
helper BS joins JT with probability $p_{b}$, which is determined
by the condition of the backhaul link between the serving BS and the
$b$-th helper BS. In the following, $p_{b}$ will be referred to
as participation probability, which can be computed from (4). Suppose
that the JT operation is scheduled to be performed at a critical time
$T$ after the serving cell pushes the UE\textquoteright{}s data into
the backhaul network. Then $p_{b}$ can be calculated as

\begin{equation}
p_{b}=\int_{0}^{T}f\left(t\right)dt=\frac{\gamma\left(\beta,\frac{T-t_{0}}{\alpha}\right)}{\Gamma\left(\beta\right)},
\end{equation}
where $\gamma\left(\cdot,\cdot\right)$ is the lower incomplete gamma
function \cite{math_book[20]}. In practice, $T$ will take a reasonably
small value to avoid performance degradation caused by outdated CSI.
For instance, if $T=10$ ms, $p_{b}\approx0.58$, and if $T=11$ ms,
$p_{b}\approx0.78$. Furthermore, if congestion occurs in the routers,
$f\left(t\right)$ will suffer from additional shift, i.e., $t_{0}$
will take large values. For instance, let $t_{0}=8.5$, then if $T=10$
ms, $p_{b}\approx0.3$, and if $T=11$ ms, $p_{b}\approx0.58$.

\section{Advances and Drawbacks of the Existing Schemes}

In this section, we discuss the advances and drawbacks of the existing
schemes, where different aspects such as system performance, required
CSI feedback, and limited backhaul connectivity are carefully examined.

\subsection{The Global Precoding Scheme}

The optimal precoding strategy for the full JT problem (3) is the
global precoding (GP), i.e., to view the distributed antenna ports
from BSs in the JT set as a giant multiple-antenna system and generalize
the well-studied point-to-point MIMO transmission strategies to JT
across multiple BSs \cite{CoMP_overview[1]}. However, the difficulty
lies in how to maintain the distributed per-BS power constraints while
extending the point-to-point MIMO schemes to network MIMO ones. To
our best knowledge, the MIN-MAX-MSE problem is still an open problem
for the linear precoding design for JT network MIMO with distributed
per-BS power constraints.

Hence, a sum power constraint is instead assumed for the GP to yield
a lower bound for the MSE performance \cite{CoMP_overview[1]}. To
formulate GP, rewrite (1) as

\begin{equation}
\mathbf{y}=\mathbf{HWx}+\mathbf{n},
\end{equation}

\noindent where $\mathbf{H}$ and $\mathbf{W}$ denote the global
channel matrix $\left[\mathbf{H}_{\mathrm{1}},\mathbf{H}_{\mathrm{2}}\right]$
and global precoder $\left[\mathbf{W}_{1}^{\textrm{T}},\mathbf{W}_{2}^{\textrm{T}}\right]^{\textrm{T}}$,
respectively. The MIN-MAX-MSE problem for GP can be re-formulated
from (3) as

\begin{equation}
\begin{array}{c}
\underset{\mathbf{F},\mathbf{W}}{\min}\quad\max\left\{ \left.M_{i}\right|i\in\left\{ 1,2,\cdots,L\right\} \right\} ,\\
\textrm{s.t.}\:\quad\textrm{tr}\left\{ \mathbf{W}^{\textrm{H}}\mathbf{W}\right\} \leq2P.\quad\quad\quad\quad
\end{array}
\end{equation}

Let $\mathbf{R}_{\mathbf{H}}=\mathbf{H}^{\textrm{H}}\mathbf{R}_{\textrm{n}}^{-1}\mathbf{H}$,
and its eigenvalue decomposition be

\begin{equation}
\mathbf{R}_{\mathbf{H}}=\mathbf{V}\mathbf{\boldsymbol{\Lambda}}\mathbf{V}^{\textrm{H}},
\end{equation}

\noindent where $\mathbf{V}\in\mathbb{C}^{2\mathit{N_{\textrm{T}}\times\mathrm{2}N_{\textrm{T}}}}$
is a unitary matrix and $\boldsymbol{\Lambda}=\textrm{diag}\left\{ \lambda_{i}\right\} $
is a semi-definite diagonal matrix, with diagonal entries $\lambda_{i}$s
being the eigenvalues of $\mathbf{R}_{\mathbf{H}}$. Then the optimal
solution for problem (7) is achieved by the joint linear transceiver
design \cite{joint_linear_transceiver[21]}. The optimal receiver
should take the form of the Wiener filter \cite{David_book[22]} shown
as

\begin{equation}
\mathbf{F}^{\textrm{opt}}=\mathbf{H}_{\textrm{eq}}^{\textrm{H}}\left(\mathbf{H}_{\textrm{eq}}\mathbf{H}_{\textrm{eq}}^{\textrm{H}}+\mathbf{R}_{\textrm{n}}\right)^{-1},
\end{equation}

\noindent where $\mathbf{H}_{\textrm{eq}}=\mathbf{HW}$ denotes the
equivalent channel. Note that the Wiener filter has been proved to
be the optimum linear receiver in the sense that it minimizes each
of the sub-stream MSEs \cite{joint_linear_transceiver[21]}. And the
optimal transmit precoding matrix $\mathbf{W}^{\textrm{opt}}$ should
be

\begin{equation}
\mathbf{W}^{\textrm{opt}}=\tilde{\mathbf{W}}\mathbf{Q}^{\textrm{H}}=\tilde{\mathbf{V}}\boldsymbol{\Sigma}\mathbf{Q}^{\textrm{H}},
\end{equation}

\noindent where $\tilde{\mathbf{W}}=\tilde{\mathbf{V}}\boldsymbol{\Sigma}$
and the column of $\tilde{\mathbf{V}}\in\mathbb{C}^{2\mathit{N_{\textrm{T}}\times L}}$
consists of the eigenvectors of $\mathbf{R}_{\mathbf{H}}$ corresponding
to the $L$ largest eigenvalues in increasing order. The power loading
matrix $\boldsymbol{\Sigma}=\mathrm{diag}\left\{ \sigma_{i}\right\} $.
The $\sigma_{i}$s are tuned so that the sum MSE with respect to $\tilde{\mathbf{W}}$
is minimized. In \cite{joint_linear_transceiver[21]}, it is proved
that the optimal $\sigma_{i}$s can be obtained by the famous water-filling
power allocation (PA) \cite{David_book[22]}

\begin{equation}
\sigma_{i}=\sqrt{\left(\mu^{-\nicefrac{1}{2}}\lambda_{i}^{-\nicefrac{1}{2}}-\lambda_{i}^{-1}\right)^{+}},
\end{equation}

\noindent where $\mu^{-\nicefrac{1}{2}}$ is the water-level chosen
to satisfy the power constraint with equality.

After minimizing the sum MSE by $\tilde{\mathbf{W}}$, a rotation
operation is applied on it. In (10), $\mathbf{Q}\in\mathbb{C}^{L\times L}$
is a unitary rotation matrix such that all sub-stream MSEs are equal.
Thereby, the minimized sum MSE is equally divided for each sub-stream,
leading to a minimized MSE $\hat{M}$ for the maximum of the sub-stream
MSEs for the problem (7), which can be concisely expressed as \cite{joint_linear_transceiver[21]},

\begin{eqnarray}
\hat{M} & = & \max\left\{ M_{i}\right\} \nonumber \\
 & = & \frac{1}{L}\textrm{tr}\left\{ \left(\mathbf{I}_{L}+\left(\mathbf{W}^{\textrm{opt}}\right)^{\textrm{H}}\mathbf{R}_{\mathbf{H}}\mathbf{W}^{\textrm{opt}}\right)^{-1}\right\} .
\end{eqnarray}

\noindent Though the closed-form expression for $\mathbf{Q}$ does
not exist, efficient algorithms to compute $\mathbf{Q}$ can be found
in \cite{Q_algo[23]}. Note that (12) is a lower bound solution for
the original problem (3) since inter-BS PA implied by (10) is usually
not the feasible solutions of (3). Besides, other practical issues
such as limited backhaul and feedback overhead also compromise the
performance of GP.

\subsection{The Autonomous Global Precoding Scheme}

\begin{algorithm*}
\begin{equation}
\mathbf{W}_{b}^{\textrm{opt}}=\begin{cases}
\mathbf{0}, & \textrm{the }b\textrm{-th BS is absent from JT},\\
\left[\mathbf{W}_{\left(b-1\right)N_{\textrm{T}}+1,:}^{\textrm{opt}\;\textrm{T}},\cdots,\mathbf{W}_{bN_{\textrm{T}},:}^{\textrm{opt}\;\textrm{T}}\right]^{\textrm{T}}, & \textrm{otherwise},
\end{cases}
\end{equation}
\end{algorithm*}

In practice, full JT, which is assumed by GP, is not always feasible
due to backhaul limitations, such as overtime delay leading to incomplete
or outdated data at the transmission points. Moreover, it is preferable
for practical systems to have distributed schedulers due to considerations
of low complexity and low cost, thereby some local scheduling constraints
in helper BSs may also force them to temporarily leave the JT set
and thus break the full JT operation. Hence, it is desirable to design
a flexible JT scheme, in which the serving BS makes the JT scheduling
decision and informs the helper BSs, then the helper BSs can adaptively
join or quit the upcoming JT operation according to their instantaneous
states. If all the helper BSs are temporarily unavailable for JT,
the system should be able to fall back to single-BS transmission (ST)
smoothly. But this gives rise to a feedback problem. The transmission
assumption, based on which the recommended precoder is computed and
fed back by the UE, may be inconsistent with the one when the transmission
eventually takes place. Consequently, the previously fed-back precoder
mismatches the transmission channel, causing performance degradation.
This problem is not uncommon, especially in 3GPP LTE-A systems \cite{LTE-A[3]}.
A straightforward solution to the above problem is to require the
UE to feedback multiple precoder recommendations under different transmission
assumptions. For example, in Fig. 1 the UE can feedback two precoders,
one for JT and another for ST.

However, aside from the issue of necessary inter-BS signaling for
switching between JT and ST precoders, additional feedback overhead
incurred from multiple precoders will become very large, since the
number of transmission assumptions can be as many as $2^{B-1}$. To
avoid increasing the feedback overhead, another approach would be
to instruct the UE to feedback the global precoder $\mathbf{W}^{\textrm{opt}}$
only. If any helper BSs are not ready for JT, they will mute themselves
during the data transmission to keep the interference low, which is
called dynamic point muting in the LTE-A system \cite{DPS_muting}.
We assume that each helper BS is unaware of the states of other helper
BSs, and thus they should stick to their respective sub-block parts
of $\mathbf{W}^{\textrm{opt}}$. Such scheme is hereafter referred
to as autonomous global precoding (AGP) and the corresponding precoder
$\mathbf{W}_{b}^{\textrm{opt}}$ for the $b$-th BS can be expressed
as (13) shown at the top of the next page. In (13), $\mathbf{W}_{i,:}$
denotes the $i$-th row of $\mathbf{W}$, thus $\left[\mathbf{W}_{\left(b-1\right)N_{\textrm{T}}+1,:}^{\textrm{opt}\;\textrm{T}},\cdots,\mathbf{W}_{bN_{\textrm{T}},:}^{\textrm{opt}\;\textrm{T}}\right]^{\textrm{T}}$
represents the sub-block part of $\mathbf{W}^{\textrm{opt}}$ spanning
from the $\left(\left(b-1\right)N_{\textrm{T}}+1\right)$-th row to
the $\left(bN_{\textrm{T}}\right)$-th row of $\mathbf{W}^{\textrm{opt}}$.
Since $\mathbf{W}^{\textrm{opt}}$ is optimized under the assumption
of full JT, its sub-block part shown in (13) may not match individual
$\mathbf{H}_{b}$ very well, which may result in large performance
degradation when the system falls back to ST or partial JT. Therefore,
in this paper we propose a flexible and adaptive precoding scheme
to alleviate the problem.

\section{The Proposed Sequential and Incremental Precoding Scheme}

We propose a sequential and incremental precoding (SIP) scheme which
is flexible and achieves satisfactory performance especially for partial
JT. In the proposed SIP scheme, the precoder optimization is performed
for each helper BS according to the descending order of participation
probabilities with the precoders of previous BSs fixed. To facilitate
the optimization, we require that the participation probabilities
be determined by the serving BS based on (5), and the descending order
of the participation probabilities be notified to the UE before it
derives the precoders. We first investigate the precoder design for
a two-BS JT network. The results are then extended to a multi-BS JT
network.

\subsection{Precoder Design for the Two-BS JT Network}

We address the optimization problem (3) for the two-BS JT network
introduced in $\textrm{Section\,\ II}$. We decouple problem (3) into
two sequential steps.

In the first step, we optimize the precoder at the serving BS to ensure
the service quality when the system falls back to ST. The problem
can be formulated as

\begin{equation}
\begin{array}{c}
\underset{\mathbf{F},\mathbf{W}_{1}}{\min}\quad\max\left\{ \left.M_{i}\right|i\in\left\{ 1,2,\cdots,L\right\} \right\} ,\\
\textrm{s.t.}\quad\;\textrm{tr}\left\{ \mathbf{W}_{1}^{\textrm{H}}\mathbf{W}_{1}\right\} \leq P.\quad\quad\quad\quad
\end{array}
\end{equation}

\noindent Note that the above problem is essentially the same as the
problem (7) except for the substitution of $\mathbf{W}$ with $\mathbf{W}_{1}$
and the maximum power. Therefore, the optimal solution for (14) can
be readily obtained as

\begin{equation}
\mathbf{W}_{1}^{\textrm{opt}}=\tilde{\mathbf{V}}_{1}\boldsymbol{\Sigma}_{1}\mathbf{Q}_{1}^{\textrm{H}},
\end{equation}

\noindent where $\tilde{\mathbf{V}}_{1}$, $\boldsymbol{\Sigma}_{1}=\mathrm{diag}\left\{ \sigma_{1,i}\right\} $
and $\mathbf{Q}_{1}$ are derived using the same method as their counterparts
in (10) without subscripts.

In the second step, we proceed to optimize the performance of the
two-BS JT with $\mathbf{W}_{1}$ fixed as $\mathbf{W}_{1}=\mathbf{W}_{1}^{\textrm{opt}}$.
This problem can be formulated as

\begin{equation}
\begin{array}{c}
\underset{\mathbf{F},\mathbf{W}_{2}}{\min}\quad\max\left\{ \left.M_{i}\right|i\in\left\{ 1,2,\cdots,L\right\} \right\} ,\\
\textrm{s.t.}\quad\;\textrm{tr}\left\{ \mathbf{W}_{2}^{\textrm{H}}\mathbf{W}_{2}\right\} \leq P.\quad\quad\quad\quad
\end{array}
\end{equation}

\noindent Problem (16) can be considered as a conditional optimization
problem for $\mathbf{W}_{2}$ with the previously derived $\mathbf{W}_{1}^{\textrm{opt}}$
fixed. Direct optimization of problem (16) is not an easy task since
the aforementioned approach of minimizing the sum MSE followed by
unitary rotation for problem (7) and (14) cannot be applied here.
Hence we resort to an iterative method to minimize the $\max\left\{ M_{i}\right\} $.

Suppose that we have a precoding matrix $\mathbf{W}_{2}^{\left(n\right)}$
in the $n$-th iteration. Then the equivalent channel can be written
as

\begin{equation}
\mathbf{H}_{\textrm{eq}}^{\left(n\right)}=\mathbf{H}_{1}\mathbf{W}_{1}^{\textrm{opt}}+\mathbf{H}_{2}\mathbf{W}_{2}^{\left(n\right)}.
\end{equation}

\noindent Similar to (9), at UE side the Wiener filter is also employed
in order to minimize the MSE, which is described as

\begin{equation}
\mathbf{F}^{\textrm{opt},\left(n\right)}=\mathbf{H}_{\textrm{eq}}^{\left(n\right)\textrm{H}}\left(\mathbf{H}_{\textrm{eq}}^{\left(n\right)}\mathbf{H}_{\textrm{eq}}^{\left(n\right)\textrm{H}}+\mathbf{R}_{\textrm{n}}\right)^{-1}.
\end{equation}

\noindent Then the MSE of the $i$-th sub-stream can be written as

\begin{equation}
M_{i}^{\left(n\right)}=\mathbb{E}\left\{ \left|\mathbf{F}_{i,:}^{\textrm{opt},\left(n\right)}\left(\mathbf{H}_{\textrm{eq}}^{\left(n\right)}\mathbf{x}+\mathbf{n}\right)-x_{i}\right|^{2}\right\} .
\end{equation}

\noindent Denote the sub-stream index associated with the maximum
sub-stream MSE as

\begin{equation}
j\left(n\right)=\underset{i}{\arg\max}\left\{ M_{i}^{\left(n\right)}\right\} .
\end{equation}

Next we update $\left(\mathbf{W}_{2}^{\left(n+1\right)}\right)_{:,j\left(n\right)}$
subject to a sub-stream power constraint $P_{2,j\left(n\right)}^{\left(n+1\right)}$
with fixed $\mathbf{F}^{\textrm{opt},\left(n\right)}$ such that $M_{j\left(n\right)}^{\left(n+1\right)}$
is minimized. Denote $\mathbf{g}=\left(\mathbf{W}_{2}^{\left(n+1\right)}\right)_{:,j\left(n\right)}$
for convenience. Then the problem (16) can be formulated as

\begin{equation}
\begin{array}{c}
\underset{\mathbf{g}}{\min}\quad M_{j\left(n\right)}^{\left(n+1\right)},\quad\quad\quad\\
\qquad\textrm{s.t.}\quad\textrm{tr}\left\{ \mathbf{g}^{\textrm{H}}\mathbf{g}\right\} \leq P_{2,j\left(n\right)}^{\left(n+1\right)},
\end{array}
\end{equation}

\noindent where $M_{j\left(n\right)}^{\left(n+1\right)}$ can be represented
in detail as (22) shown at the top of the next page.
\begin{algorithm*}
\begin{eqnarray}
M_{j\left(n\right)}^{\left(n+1\right)} & = & \mathbb{E}\left\{ \left|\mathbf{F}_{j\left(n\right),:}^{\textrm{opt},\left(n\right)}\left(\left(\mathbf{H}_{1}\mathbf{W}_{1}^{\textrm{opt}}+\mathbf{H}_{2}\mathbf{W}_{2}^{\left(n+1\right)}\right)\mathbf{x}+\mathbf{n}\right)-x_{j\left(n\right)}\right|^{2}\right\} \nonumber \\
 & = & \sum_{i\neq j\left(n\right)}\left|\mathbf{F}_{j\left(n\right),:}^{\textrm{opt},\left(n\right)}\left(\mathbf{H}_{1}\left(\mathbf{W}_{1}^{\textrm{opt}}\right)_{:,i}+\mathbf{H}_{2}\left(\mathbf{W}_{2}^{\left(n+1\right)}\right)_{:,i}\right)\right|^{2}\nonumber \\
 &  & +\left|\mathbf{F}_{j\left(n\right),:}^{\textrm{opt},\left(n\right)}\left(\mathbf{H}_{1}\left(\mathbf{W}_{1}^{\textrm{opt}}\right)_{:,j\left(n\right)}+\mathbf{H}_{2}\mathbf{g}\right)\right|^{2}\nonumber \\
 &  & -2\textrm{Re}\left\{ \mathbf{F}_{j\left(n\right),:}^{\textrm{opt},\left(n\right)}\left(\mathbf{H}_{1}\left(\mathbf{W}_{1}^{\textrm{opt}}\right)_{:,j\left(n\right)}+\mathbf{H}_{2}\mathbf{g}\right)\right\} \nonumber \\
 &  & +\mathbf{F}_{j\left(n\right),:}^{\textrm{opt},\left(n\right)}\mathbf{R}_{\mathrm{n}}\left(\mathbf{F}_{j\left(n\right),:}^{\textrm{opt},\left(n\right)}\right)^{\textrm{H}}+N_{0}.
\end{eqnarray}
\end{algorithm*}
We further fix $\left(\mathbf{W}_{2}^{\left(n+1\right)}\right)_{:,i}$
for $i\neq j\left(n\right)$. Omitting the irrelevant terms in (22)
for simplicity, we get

\vspace{0.5em}

\noindent $\tilde{M}_{j\left(n\right)}^{\left(n+1\right)}=\left|\mathbf{F}_{j\left(n\right),:}^{\textrm{opt},\left(n\right)}\left(\mathbf{H}_{1}\left(\mathbf{W}_{1}^{\textrm{opt}}\right)_{:,j\left(n\right)}+\mathbf{H}_{2}\mathbf{g}\right)\right|^{2}$

\begin{equation}
-2\textrm{Re}\left\{ \mathbf{F}_{j\left(n\right),:}^{\textrm{opt},\left(n\right)}\left(\mathbf{H}_{1}\left(\mathbf{W}_{1}^{\textrm{opt}}\right)_{:,j\left(n\right)}+\mathbf{H}_{2}\mathbf{g}\right)\right\} .
\end{equation}

\noindent Hence, problem (21) is equivalent to the following problem

\begin{equation}
\begin{array}{c}
\underset{\mathbf{g}}{\min}\quad\tilde{M}_{j\left(n\right)}^{\left(n+1\right)},\quad\quad\quad\\
\qquad\textrm{s.t.}\quad\textrm{tr}\left\{ \mathbf{g}^{\textrm{H}}\mathbf{g}\right\} \leq P_{2,j\left(n\right)}^{\left(n+1\right)}.
\end{array}
\end{equation}

\noindent It is easy to verify that the problem (24) is convex. Thus,
we can obtain the optimal $\mathbf{g}$ from the KKT conditions \cite{CVX[24]}.
The Lagrangian function of (24) is given by

\begin{equation}
\mathcal{L}\left(\mathbf{g},\eta\right)=\tilde{M}_{j\left(n\right)}^{\left(n+1\right)}+\eta\left(\textrm{tr}\left\{ \mathbf{g}^{\textrm{H}}\mathbf{g}\right\} -P_{2,j\left(n\right)}^{\left(n+1\right)}\right),
\end{equation}

\noindent where $\eta\geq0$ is the Lagrangian multiplier. Taking
its derivative with respect to $\mathbf{g}^{\textrm{*}}$, we have

\vspace{0.5em}

\noindent $\frac{\partial\mathcal{L}}{\partial\mathbf{g}^{\textrm{*}}}=\mathbf{H}_{2}^{\textrm{H}}\left(\mathbf{F}_{j\left(n\right),:}^{\textrm{opt},\left(n\right)}\right)^{\textrm{H}}\mathbf{F}_{j\left(n\right),:}^{\textrm{opt},\left(n\right)}\left(\mathbf{H}_{1}\left(\mathbf{W}_{1}^{\textrm{opt}}\right)_{:,j\left(n\right)}+\mathbf{H}_{2}\mathbf{g}\right)$

\begin{equation}
-\mathbf{H}_{2}^{\textrm{H}}\left(\mathbf{F}_{j\left(n\right),:}^{\textrm{opt},\left(n\right)}\right)^{\textrm{H}}+\eta\mathbf{g}.
\end{equation}

\noindent The KKT conditions are as follows

\begin{equation}
\begin{cases}
\frac{\partial\mathcal{L}}{\partial\mathbf{g}^{\textrm{*}}}=0, & (\textrm{a})\\
\eta\left(\textrm{tr}\left\{ \mathbf{g}^{\textrm{H}}\mathbf{g}\right\} -P_{2,j\left(n\right)}^{\left(n+1\right)}\right)=0, & (\textrm{b})\\
\textrm{tr}\left\{ \mathbf{g}^{\textrm{H}}\mathbf{g}\right\} \leq P_{2,j\left(n\right)}^{\left(n+1\right)}. & (\textrm{c})
\end{cases}
\end{equation}

\noindent From (27), the closed-form expression for $\mathbf{g}$
can be derived as

\vspace{0.5em}

\noindent $\mathbf{g}=\left(\mathbf{H}_{2}^{\textrm{H}}\left(\mathbf{F}_{j\left(n\right),:}^{\textrm{opt},\left(n\right)}\right)^{\textrm{H}}\mathbf{F}_{j\left(n\right),:}^{\textrm{opt},\left(n\right)}\mathbf{H}_{2}+\eta\mathbf{I}\right)^{-1}$

\begin{equation}
\times\mathbf{H}_{2}^{\textrm{H}}\left(\left(\mathbf{F}_{j\left(n\right),:}^{\textrm{opt},\left(n\right)}\right)^{\textrm{H}}-\left(\mathbf{F}_{j\left(n\right),:}^{\textrm{opt},\left(n\right)}\right)^{\textrm{H}}\mathbf{F}_{j\left(n\right),:}^{\textrm{opt},\left(n\right)}\mathbf{H}_{1}\left(\mathbf{W}_{1}^{\textrm{opt}}\right)_{:,j\left(n\right)}\right),
\end{equation}

\noindent where $\eta$ should be chosen such that (27.b) and (27.c)
are satisfied.

The remaining problem is the power allocation (PA) strategy for $P_{2,j\left(n\right)}^{\left(n+1\right)}$.
As shown by \cite{joint_linear_transceiver[21]}, the optimal solution
for the MIN-MAX-MSE problem (7) is achieved when the sub-stream MSEs
are equal. Motivated by this result, we propose to transfer a small
amount of power from the sub-stream with minimum MSE to that with
maximum MSE in each iteration. In such way, the maximum MSE $M_{j\left(n\right)}^{\left(n\right)}$
in the $n$-th iteration will decrease to $M_{j\left(n\right)}^{\left(n+1\right)}$
due to the optimized precoding vector $\mathbf{g}$ together with
additional power bonus received from the sub-stream with minimum MSE.

Let

\noindent
\begin{equation}
k\left(n\right)=\underset{i}{\arg\min}\left\{ M_{i}^{\left(n\right)}\right\} ,
\end{equation}
we update the PA as follows

\begin{equation}
\begin{cases}
P_{2,j\left(n\right)}^{\left(n+1\right)}=P_{2,j\left(n\right)}^{\left(n\right)}+\delta P_{2,k\left(n\right)}^{\left(n\right)},\\
P_{2,k\left(n\right)}^{\left(n+1\right)}=P_{2,k\left(n\right)}^{\left(n\right)}\times\left(1-\delta\right),\\
P_{2,i}^{\left(n+1\right)}=P_{2,i}^{\left(n\right)}, & \textrm{for }i\neq j\left(n\right),k\left(n\right),
\end{cases}
\end{equation}

\noindent where

\noindent
\begin{equation}
P_{2,l}^{\left(n\right)}=\left|\left(\mathbf{W}_{2}^{\left(n\right)}\right)_{:,l}\right|^{2},l\in\left\{ 1,2,\cdots,L\right\} ,
\end{equation}
and $\delta$ is the percentage of power transferred from the $k\left(n\right)$-th
sub-stream to the $j\left(n\right)$-th sub-stream. Based on (28)
and (30), $\mathbf{W}_{2}^{\left(n+1\right)}$ can be updated as

\begin{equation}
\begin{cases}
\left(\mathbf{W}_{2}^{\left(n+1\right)}\right)_{:,j\left(n\right)}=\mathbf{g},\\
\left(\mathbf{W}_{2}^{\left(n+1\right)}\right)_{:,k\left(n\right)}=\sqrt{1-\delta}\left(\mathbf{W}_{2}^{\left(n\right)}\right)_{:,k\left(n\right)},\\
\left(\mathbf{W}_{2}^{\left(n+1\right)}\right)_{:,i}=\left(\mathbf{W}_{2}^{\left(n\right)}\right)_{:,i},\:\textrm{for }i\neq j\left(n\right),k\left(n\right).
\end{cases}
\end{equation}
Note that the power allocation should be initialized such that $\sum_{l=1}^{L}P_{2,l}^{\left(0\right)}=P$
so that in the following iterations, the power constraint of $\mathbf{W}_{2}^{\left(n\right)}$
can always be satisfied. When the iterative algorithm converges, all
sub-stream MSEs should be equal. Hence, the termination criterion
can be established based on the difference between $M_{j\left(n\right)}^{\left(n\right)}$
and $M_{k\left(n\right)}^{\left(n\right)}$. The proposed precoding
scheme will be referred to as the sequential and incremental precoding
(SIP) scheme hereafter and is summarized in Algorithm 1.

\textit{}
\begin{algorithm}
\noindent \textit{\caption{\textit{The SIP Scheme for Two-BS JT}}
}\textit{\emph{}}\\
\textit{\emph{Step 1:}}\textit{ }Compute $\mathbf{W}_{1}^{\textrm{opt}}$
according to (15);

\noindent \textit{\emph{Step 2:}}\textit{ }Obtain $\mathbf{W}_{2}^{\textrm{opt}}$
\textit{\emph{using the following iterative algorithm,}}
\begin{enumerate}
\item Initialization: \\
Set $\mathbf{W}_{2}^{\left(1\right)}=\sqrt{\frac{P}{L}}\left[\mathbf{I}_{L},\boldsymbol{0}_{L\times\left(N_{\textrm{T}}-L\right)}\right]^{\textrm{T}}$
and $n=1$.
\item Iteration: \\
(a) Compute $\mathbf{H}_{\textrm{eq}}^{\left(n\right)}$ and $\mathbf{F}^{\textrm{opt},\left(n\right)}$
using (17) and (18), respectively; \\
(b) Use (28) to compute $\mathbf{g}$ and (32) to get $\mathbf{W}_{2}^{\left(n+1\right)}$;
\item Termination: \\
The algorithm terminates either when $M_{j\left(n\right)}^{\left(n\right)}$
and $M_{k\left(n\right)}^{\left(n\right)}$ converges, i.e., $\frac{\left|M_{j\left(n\right)}^{\left(n\right)}-M_{k\left(n\right)}^{\left(n\right)}\right|}{M_{j\left(n\right)}^{\left(n\right)}}\leq\xi_{\textrm{th}}$
or when $n\geq N_{\textrm{max}}$, where $\xi_{\textrm{th}}$ is a
predefined threshold and $N_{\textrm{max}}$ is the maximum iteration
number; \\
Output $\mathbf{W}_{2}^{\textrm{opt}}=\mathbf{W}_{2}^{\left(n\right)}$.\\
Else, $n=n+1$, and go to sub-step 2). \end{enumerate}
\end{algorithm}

\subsection{Extension of the SIP Scheme}

In this subsection, we generalize our proposed SIP scheme to multi-BS
scenarios, where $B>2$. Without loss of generality, the $B-1$ helper
BSs are sorted by their participation probabilities in decreasing
order as: $p_{2}\geq p_{3}\geq\cdots\geq p_{B}$. Then the corresponding
precoders are sequentially and incrementally optimized with $\mathbf{W}_{2}$
first and $\mathbf{W}_{B}$ last based on Step 2 of the proposed SIP
scheme. To be more specific, $\mathbf{W}_{b}\left(b\in\left\{ 2,3,\cdots,B\right\} \right)$
is optimized with fixed $\mathbf{W}_{i}^{\textrm{opt}}\left(i\in\left\{ 1,2,\cdots,b-1\right\} \right)$
previously obtained from the SIP scheme. In addition, the equivalent
channel shown in (17) now should be computed as

\begin{equation}
\mathbf{H}_{\textrm{eq}}^{\left(n\right)}=\sum_{i=1}^{b-1}\mathbf{H}_{i}\mathbf{W}_{i}^{\textrm{opt}}+\mathbf{H}_{b}\mathbf{W}_{b}^{\left(n\right)}.
\end{equation}

\noindent If helper BS $b$ joins JT, its precoder will be $\mathbf{W}_{b}^{\textrm{opt}}$.
Otherwise, it mutes its transmission.

To sum up, the proposed SIP scheme can be extended to be employed
in a multi-BS JT scenario and is summarized in Algorithm 2.

\textit{}
\begin{algorithm}
\noindent \textit{\caption{\textit{The SIP Scheme for Multi-BS JT}}
}\textit{\emph{}}\\
\textit{\emph{Step 1:}}\textit{ }Compute $\mathbf{W}_{1}^{\textrm{opt}}$
according to (15);

\noindent \textit{\emph{~~~~~~~~~Set}}\textit{ }$b=2$;

\noindent \textit{\emph{Step 2:}}\textit{ }Obtain $\mathbf{W}_{b}^{\textrm{opt}}$
\textit{\emph{using the following iterative algorithm,}}
\begin{enumerate}
\item Initialization: \\
Set $\mathbf{W}_{b}^{\left(1\right)}=\sqrt{\frac{P}{L}}\left[\mathbf{I}_{L},\boldsymbol{0}_{L\times\left(N_{\textrm{T}}-L\right)}\right]^{\textrm{T}}$
and $n=1$;
\item Iteration: \\
(a) Compute $\mathbf{H}_{\textrm{eq}}^{\left(n\right)}$ and $\mathbf{F}^{\textrm{opt},\left(n\right)}$
using (33) and (18), respectively; \\
(b) Use (28) to compute $\mathbf{g}$ and (32) to get $\mathbf{W}_{b}^{\left(n+1\right)}$;
\item Termination: \\
The algorithm terminates either when $M_{j\left(n\right)}^{\left(n\right)}$
and $M_{k\left(n\right)}^{\left(n\right)}$ converges, i.e., $\frac{\left|M_{j\left(n\right)}^{\left(n\right)}-M_{k\left(n\right)}^{\left(n\right)}\right|}{M_{j\left(n\right)}^{\left(n\right)}}\leq\xi_{\textrm{th}}$
or when $n\geq N_{\textrm{max}}$, where $\xi_{\textrm{th}}$ is a
predefined threshold and $N_{\textrm{max}}$ is the maximum iteration
number; \\
Output $\mathbf{W}_{b}^{\textrm{opt}}=\mathbf{W}_{b}^{\left(n\right)}$;\\
Else, $n=n+1$, then go to sub-step 2).
\end{enumerate}
\noindent \textit{\emph{Step 3: If}}\textit{ }$b<B$, then $b=b+1$,
and go to Step 2

\noindent \textit{\emph{Else, terminate the algorithm with }}$\mathbf{W}_{b}^{\textrm{opt}}$$\left(b\in\left\{ 1,2,\cdots,B\right\} \right)$
as the per-BS precoders.
\end{algorithm}

\subsection{The SIP Scheme with Codebook Based Feedback }

If a codebook, denoted as $\boldsymbol{\Omega}$, is employed as the
set of precoder candidates, then UE can exhaustively search $\boldsymbol{\Omega}$,
find the best precoder and feedback its index to the BS using just
a few bits. In practice, codebook based feedback is commonly used
in FDD systems such as the LTE-A system \cite{LTE-A[3]} due to low
overhead costs. Here, we pursue the philosophy of sequential and incremental
precoding, and propose the SIP scheme with codebook based feedback.

For $\mathbf{W}_{1}$, the best precoder in the codebook $\boldsymbol{\Omega}$
can be written as

\begin{equation}
\mathbf{W}_{1}^{\textrm{opt,cb}}=\underset{\mathbf{W}_{1}^{\textrm{cb}}\in\boldsymbol{\Omega}}{\arg\min}\;\max\left\{ \left.M_{i}\right|i\in\left\{ 1,2,\cdots,L\right\} \right\} ,
\end{equation}
where $M_{i}=\mathbb{E}\left\{ \left|\mathbf{F}_{i,:}^{\textrm{opt}}\left(\mathbf{H}_{1}\mathbf{W}_{1}^{\textrm{cb}}\mathbf{x}+\mathbf{n}\right)-x_{i}\right|^{2}\right\} $.

For $\mathbf{W}_{b}$, we fix the previously optimized precoders and
incrementally find the best precoder for $\mathbf{W}_{b}$ from

\begin{equation}
\mathbf{W}_{b}^{\textrm{opt,cb}}=\underset{\mathbf{W}_{b}^{\textrm{cb}}\in\boldsymbol{\Omega}}{\arg\min}\!\max\left\{ \left.M_{i}\right|i\in\left\{ 1,2,\cdots,L\right\} \right\} ,
\end{equation}
where $M_{i}=\mathbb{E}\left\{ \left|\mathbf{F}_{i,:}^{\textrm{opt}}\left(\mathbf{H}_{\textrm{eq}}\mathbf{x}+\mathbf{n}\right)-x_{i}\right|^{2}\right\} $
and $\mathbf{H}_{\textrm{eq}}=\sum_{i=1}^{b-1}\mathbf{H}_{i}\mathbf{W}_{i}^{\textrm{opt,cb}}+\mathbf{H}_{b}\mathbf{W}_{b}^{\textrm{cb}}$.

In the AGP scheme, the best precoder from the global codebook $\boldsymbol{\Omega}_{\textrm{GP}}$
can be represented as

\begin{equation}
\mathbf{W}^{\textrm{opt,cb}}=\underset{\mathbf{W}^{\textrm{cb}}\in\boldsymbol{\Omega}_{\textrm{GP}}}{\arg\min}\!\max\left\{ \left.M_{i}\right|i\in\left\{ 1,2,\cdots,L\right\} \right\} ,
\end{equation}
where $M_{i}=\mathbb{E}\left\{ \left|\mathbf{F}_{i,:}^{\textrm{opt}}\left(\mathbf{H}\mathbf{W}^{\textrm{cb}}\mathbf{x}+\mathbf{n}\right)-x_{i}\right|^{2}\right\} $.
Then the $\mathbf{W}_{b}^{\textrm{opt,cb}}$s of the AGP scheme can
be readily obtained from (13) with $\mathbf{W}^{\textrm{opt}}$ replaced
by $\mathbf{W}^{\textrm{opt,cb}}$.

Suppose that the cardinality of $\boldsymbol{\Omega}$ is $2^{d}$,
then for each BS $d$ bits are needed to feedback $\mathbf{W}_{b}^{\textrm{opt,cb}}$
from $2^{d}$ precoder candidates. In order to make a fair comparison
between the SIP and AGP schemes, the cardinality of $\boldsymbol{\Omega}_{\textrm{GP}}$
should be $2^{Bd}$, i.e., a total overhead of $Bd$ bits are assumed
for the feedback of precoders in both schemes.

\section{Simulation Results and Discussions}

In this section, we present simulation results to compare the maximum
MSE and average BER performances of the proposed SIP scheme with those
of the AGP scheme. We consider a practical setup where a single multi-antenna
UE with $N_{\textrm{R}}=2$ or 4 is served by a multi-BS JT set with
$B=2$ or 3 and $N_{\textrm{T}}=4$. Suppose that the critical time
$T=11$ ms, and $t_{0}$ for $\mathrm{BS}_{2}$ and $\mathrm{BS}_{3}$
are set to $t_{0,2}=7.5$ ms and $t_{0,3}=8.5$ ms, respectively.
As explained in Section II-B, the corresponding participation probabilities
for $\mathrm{BS}_{2}$ and $\mathrm{BS}_{3}$ can be calculated using
(5) and we get $p_{2}\approx0.78$ and $p_{3}\approx0.58$. Note that
the water filling PA for the AGP and SIP schemes may lead to rank
adaptation, i.e. dropping sub-streams with poor channel gains. For
fairness, rank adaptation should be forbidden in our simulations.
Therefore, when $N_{\textrm{R}}=4$ we fix $L=4$ to prevent rank
adaptation and apply equal sub-stream PA to $\mathbf{W}_{1}$, i.e.,
set $\sigma_{1,i}=P/L$ in (15), whereas the power transfer shown
in (30) is applied for $\mathbf{W}_{b}$s when $b\neq1$. The water
filling PA is only engaged for $\mathbf{W}_{1}$ in the case of $N_{\textrm{R}}=2$,
when rank adaptation rarely happens. At the UE side, we assume that
the Wiener filter is always employed.

We define per-BS signal to interference plus noise ratio (SINR) by
$SINR=P/N_{0}$. All channels are assumed to experience uncorrelated
Rayleigh fading and the entries of $\mathbf{H}_{b}$ are i.i.d. zero-mean
circularly symmetric complex Gaussian (ZMCSCG) random variables with
unit variance. The results are averaged over 10, 000 independent channel
realizations. As for the BER results, 1, 000, 000 symbols obtained
from the QPSK constellation are transmitted in each channel realization
for each simulated SINR point. In addition, for the proposed SIP scheme,
we set the convergence threshold $\xi_{\textrm{th}}=0.01$, power
transfer percentage $\delta=1\%$ and the maximum iteration number
$N_{\textrm{max}}=100$.

\subsection{Convergence of SIP Scheme}

Before discussing the numerical results of the system performance,
we first investigate the convergence behavior of the proposed SIP
scheme summarized in algorithm 1. Fig. 4 and 5 show the mean of the
maximum of sub-stream MSEs versus number of iterations for $B=2$
and $N_{\textrm{R}}=2$ or 4 with different $SINR$. As seen from
these two figures, the MSE always converges. When $N_{\textrm{R}}=2$,
the MSE converges typically after 20 iterations, and more iterations
are needed for the case of $N_{\textrm{R}}=4$. Moreover, the convergence
of the SIP scheme for multi-BS JT is straightforward since the mathematical
form of $\mathbf{H}_{\textrm{eq}}^{\left(n\right)}$ in algorithm
2 is essentially the same as that in algorithm 1. Therefore, here
we omit the illustration of algorithm convergence for the case of
$B=3$.

\begin{figure}[H]
\centering\includegraphics[width=8cm]{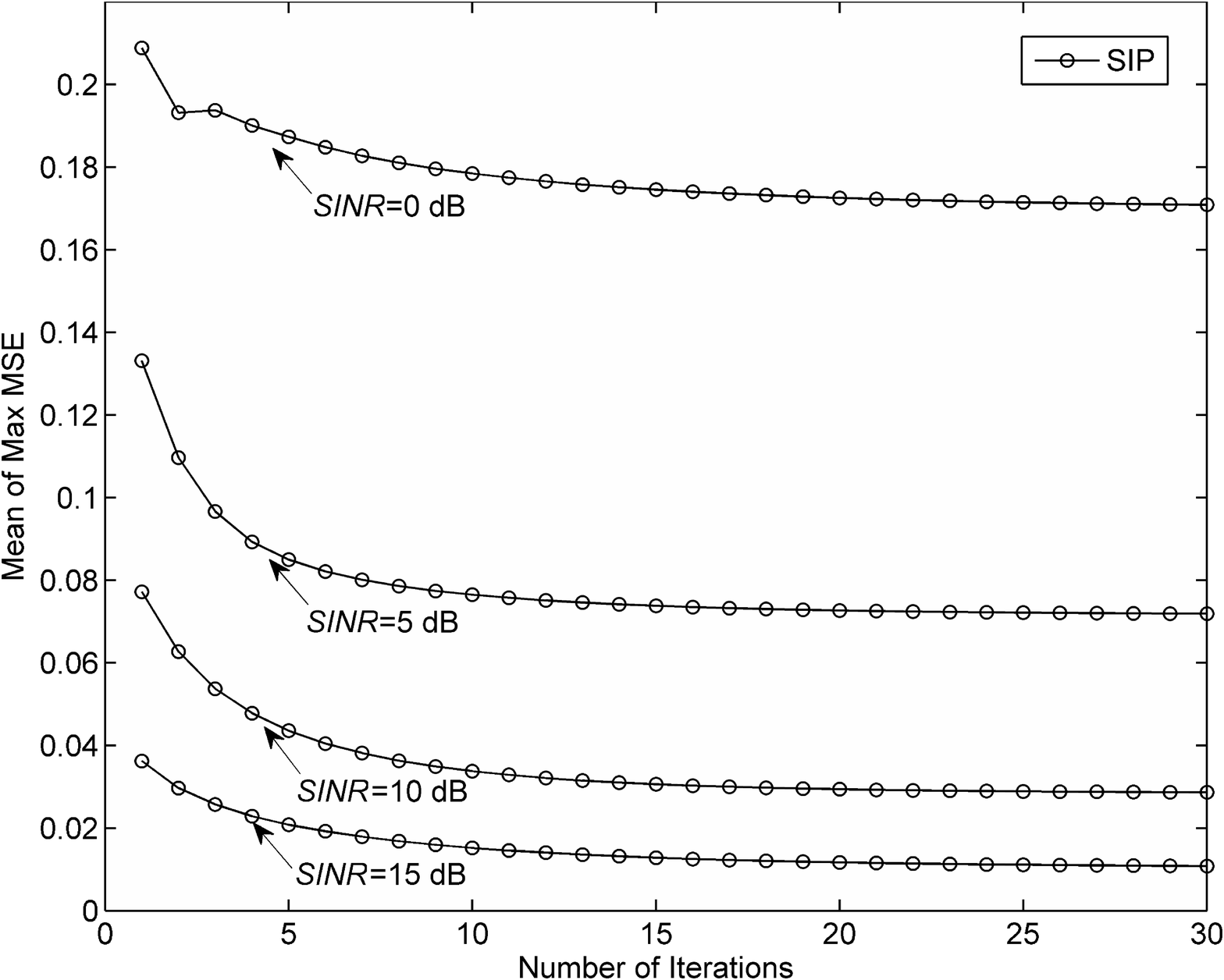} 
\vspace{-0.5em}
\renewcommand{\figurename}{Fig.}\caption{Convergence of the SIP scheme for $B=2$, $N_{\mathrm{R}}=2$ (perfect
feedback).}
\end{figure}
\begin{figure}[H]
\centering\includegraphics[width=8cm]{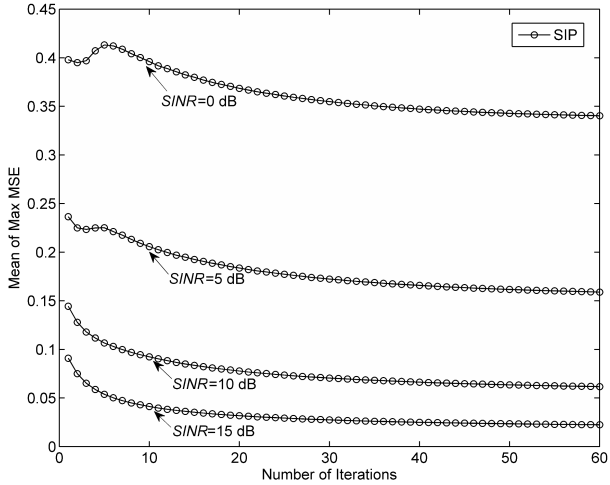} 
\vspace{-0.5em}
\renewcommand{\figurename}{Fig.}\caption{Convergence of the SIP scheme for $B=2$, $N_{\mathrm{R}}=4$ (perfect
feedback).}
\end{figure}

\subsection{Performance of the Mean of the Maximum of Sub-stream}

Fig. 6 and 7 show the average performance of the maximum of sub-stream
MSEs for $B=2$ and $N_{\textrm{R}}=2$ or 4 with different $p_{2}$.
For the case of $p_{2}=0$, the system degenerates to ST due to broken
backhaul, while the case of $p_{2}=1$ corresponds to full JT with
perfect backhaul. As explained in Section III and observed in Fig.
6 and 7, the precoder for the AGP scheme is optimized under the assumption
of full JT, which incurs large performance degradation when the system
falls back to ST or partial JT. When the practical backhaul is considered,
i.e., $p_{2}=0.78$, the proposed SIP scheme offers significant performance
gain and the gain is more pronounced in high SINR regimes because
the sequentially and incrementally designed precoder matches the actual
transmission channel better than the precoder in AGP which is optimized
for full JT.

\begin{figure}[H]
\centering\includegraphics[width=8cm]{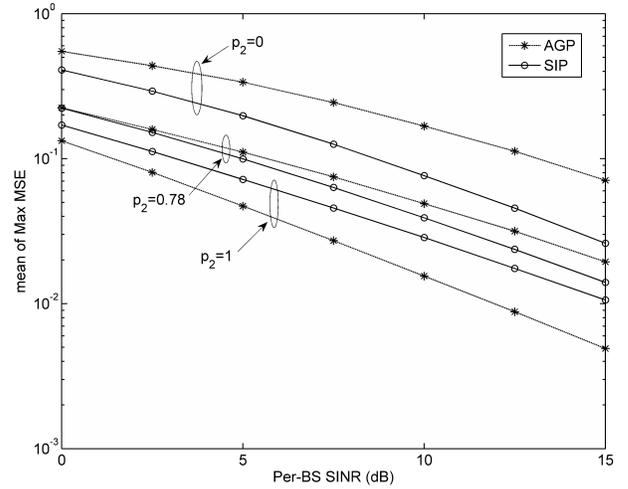} 
\vspace{-0.5em}
\renewcommand{\figurename}{Fig.}\caption{Mean of the maximum of sub-stream MSEs for $B=2$, $N_{\mathrm{R}}=2$
(perfect feedback).}
\end{figure}
\begin{figure}[H]
\centering\includegraphics[width=8cm]{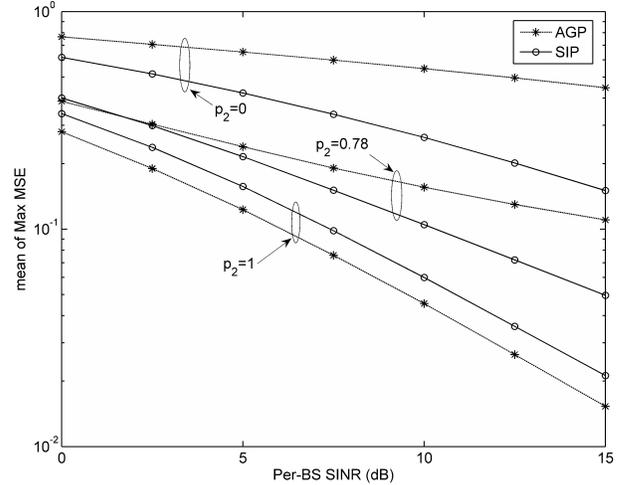} 
\vspace{-0.5em}
\renewcommand{\figurename}{Fig.}\caption{Mean of the maximum of sub-stream MSEs for $B=2$, $N_{\mathrm{R}}=4$
(perfect feedback).}
\end{figure}

\subsection{Average BER Performance}

Fig. 8 and 9 show the average BER performance for $B=2$ and $N_{\textrm{R}}=2$
or 4 with different $p_{2}$. As seen from Fig. 8 and 9, our proposed
SIP scheme also shows superior BER performance when $p_{2}=0.78$,
especially in high SINR regimes. When $SINR=15$ dB, compared with
the AGP scheme, the proposed SIP scheme can reduce the average BER
from $0.5\times10^{-3}$ to $10^{-5}$ and from $10^{-2}$ to $3\times10^{-3}$
for the case of $N_{\textrm{R}}=2$ and $N_{\textrm{R}}=4$, respectively.

\begin{figure}[H]
\centering\includegraphics[width=8cm]{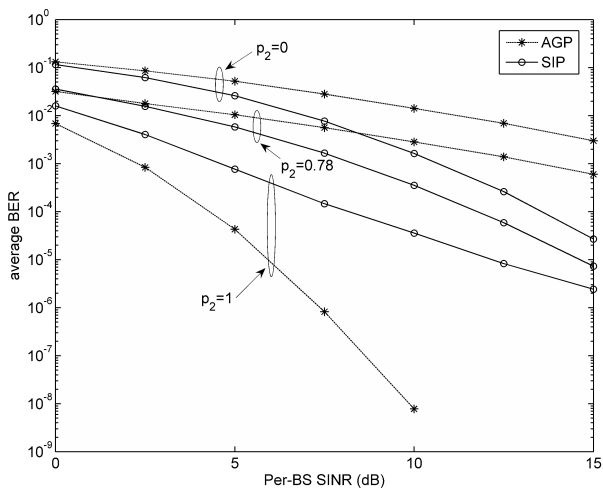} 
\vspace{-0.5em}
\renewcommand{\figurename}{Fig.}\caption{Average BER for $B=2$, $N_{\mathrm{R}}=2$ (perfect feedback).}
\end{figure}
\begin{figure}[H]
\centering\includegraphics[width=8cm]{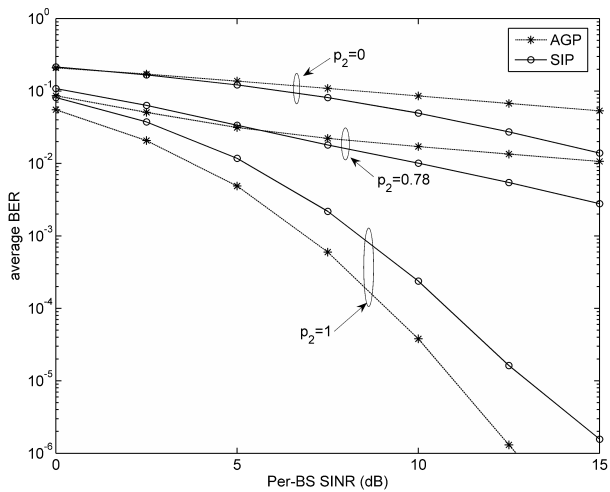} 
\vspace{-0.5em}
\renewcommand{\figurename}{Fig.}\caption{Average BER for $B=2$, $N_{\mathrm{R}}=4$ (perfect feedback).}
\end{figure}

\subsection{Performance of the Extended SIP Scheme}

For the extended case of $B=3$, we show the average performance of
the maximum of sub-stream MSEs in Fig. 10 and 11, and the corresponding
BER results in Fig. 12 and 13 for $N_{\textrm{R}}=2$ or 4 with different
$(p_{2},p_{3})$. Much like what we have observed previously, when
the backhaul suffers from limited connectivity, e.g., $(p_{2},p_{3})=(0.78,\,0.58)$,
the proposed SIP scheme significantly outperforms the AGP scheme in
terms of average BER when SINR is high. It is very interesting to
note that the SIP and AGP schemes exhibit close averaged maximum MSE
curves when $(p_{2},p_{3})=(0.78,\,0.58)$ in Fig. 10, but they have
notable BER difference in favor of the SIP scheme in Fig. 12. One
possible explanation might be that the overall MSE is also important
to determine the BER performance and the average MSE of the AGP scheme
may not be well-controlled as the maximum MSE. To investigate this
issue, we conduct simulations to show the average MSE performance
in Fig. 14 using parameters given in Fig. 10. From Fig. 14, we observe
that the proposed SIP scheme doesn't outperform the AGP scheme in
terms of average MSE, because the SIP scheme targets the optimization
of the maximum sub-stream MSE shown in problem (16), not the average
MSE. Hence, the comparison of the average MSE performance does not
highly relate to that of the BER performance. We suppose that the
reason is that although the average maximum MSE performance is similar,
the maximum sub-stream MSE of the AGP scheme varies more widely than
that of the SIP scheme, as can be observed from Fig. 10 where the
variant range of the maximum sub-stream MSE is roughly bounded by
the \textquotedblleft{}curves of mean of Max MSE\textquotedblright{}
for $(p_{2},p_{3})=(0,\,0)$ and $(p_{2},p_{3})=(1,\,1)$, and obviously
the variant range of SIP is narrower than that of AGP. The variant
range of the maximum sub-stream MSE indicates that the AGP scheme
tends to generate larger maximum sub-stream MSE than the SIP scheme
does in some poor cases, and the average BER performance is dominated
by the large BERs resulted from poor-case maximum sub-stream MSEs,
which will lead to a higher BER performance for the AGP scheme.

\begin{figure}[H]
\centering\includegraphics[width=8cm]{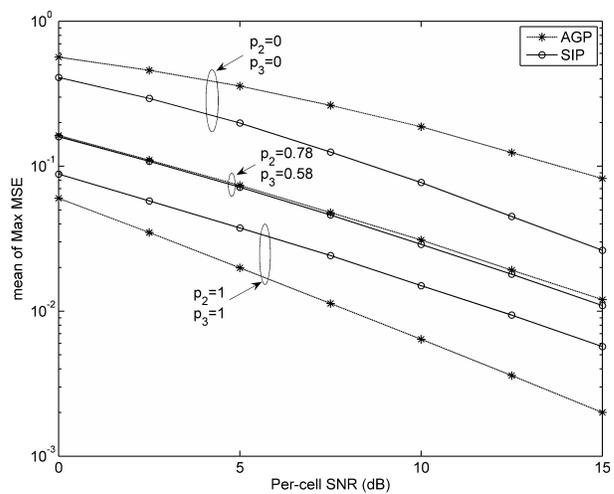} 
\vspace{-0.5em}
\renewcommand{\figurename}{Fig.}\caption{Mean of the maximum of sub-stream MSEs for $B=3$, $N_{\mathrm{R}}=2$
(perfect feedback).}
\end{figure}
\begin{figure}[H]
\centering\includegraphics[width=8cm]{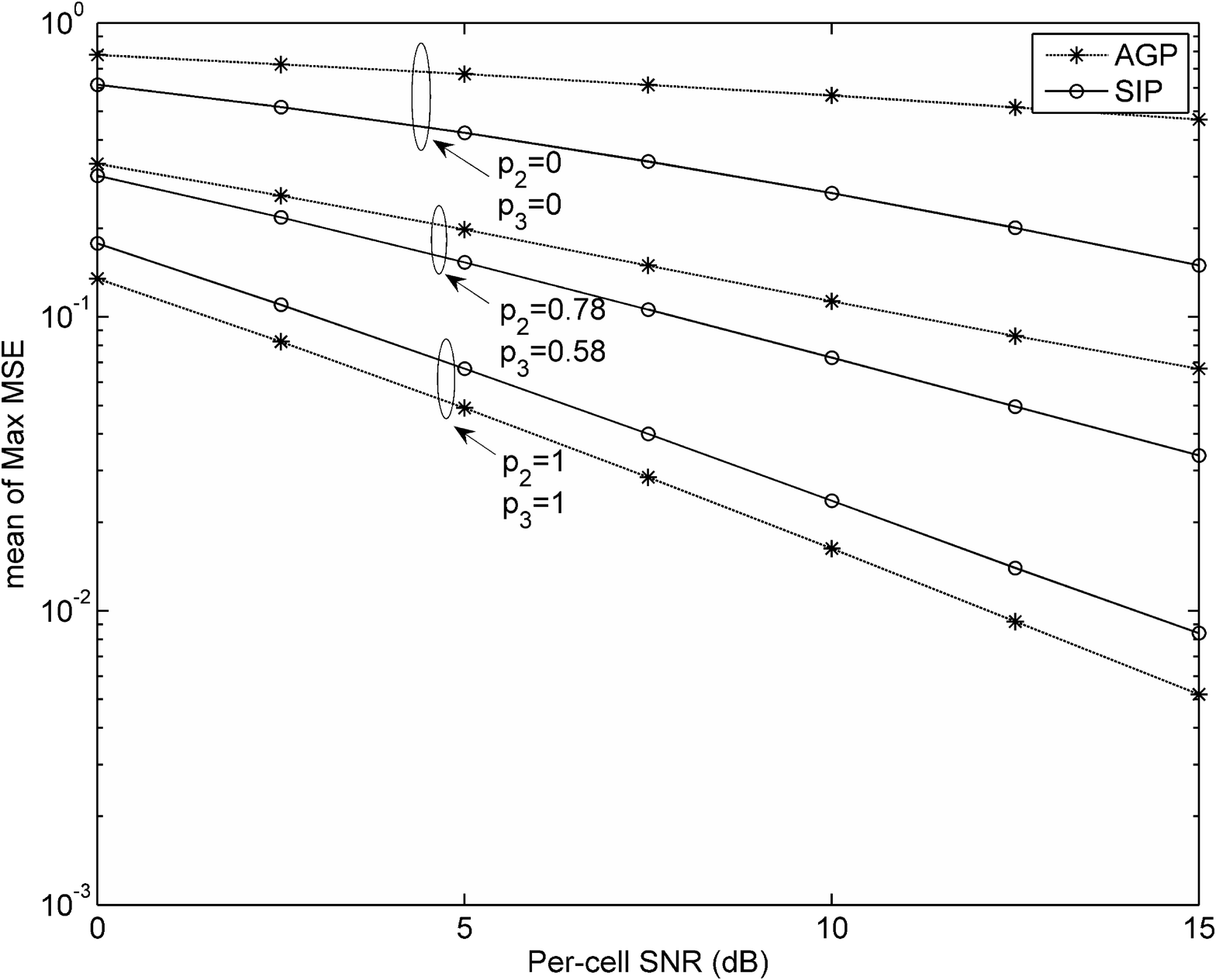} 
\vspace{-0.5em}
\renewcommand{\figurename}{Fig.}\caption{Mean of the maximum of sub-stream MSEs for $B=3$, $N_{\mathrm{R}}=4$
(perfect feedback).}
\end{figure}
\begin{figure}[H]
\centering\includegraphics[width=8cm]{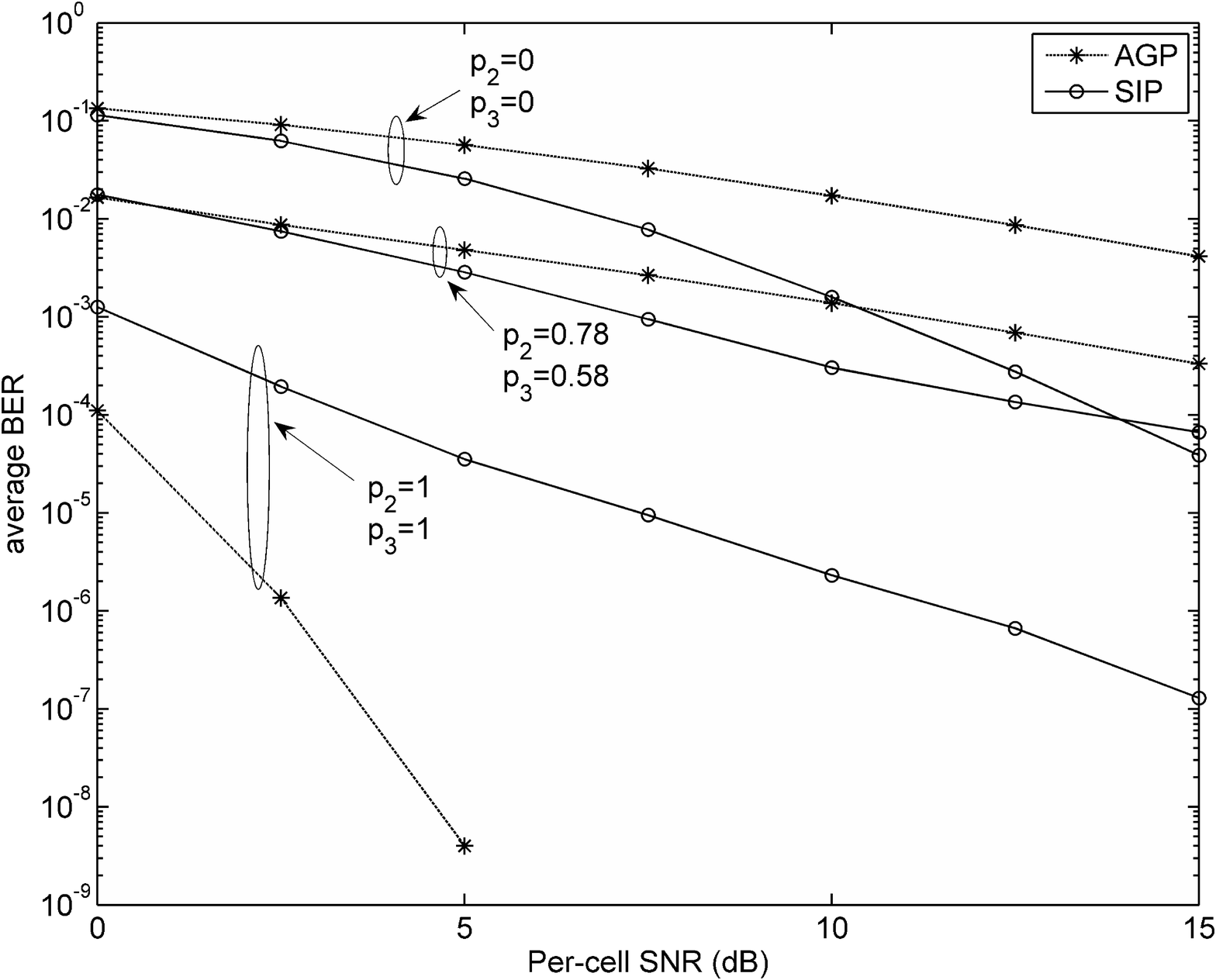} 
\vspace{-0.5em}
\renewcommand{\figurename}{Fig.}\caption{Average BER for $B=3$, $N_{\mathrm{R}}=2$ (perfect feedback).}
\end{figure}
\begin{figure}[H]
\centering\includegraphics[width=8cm]{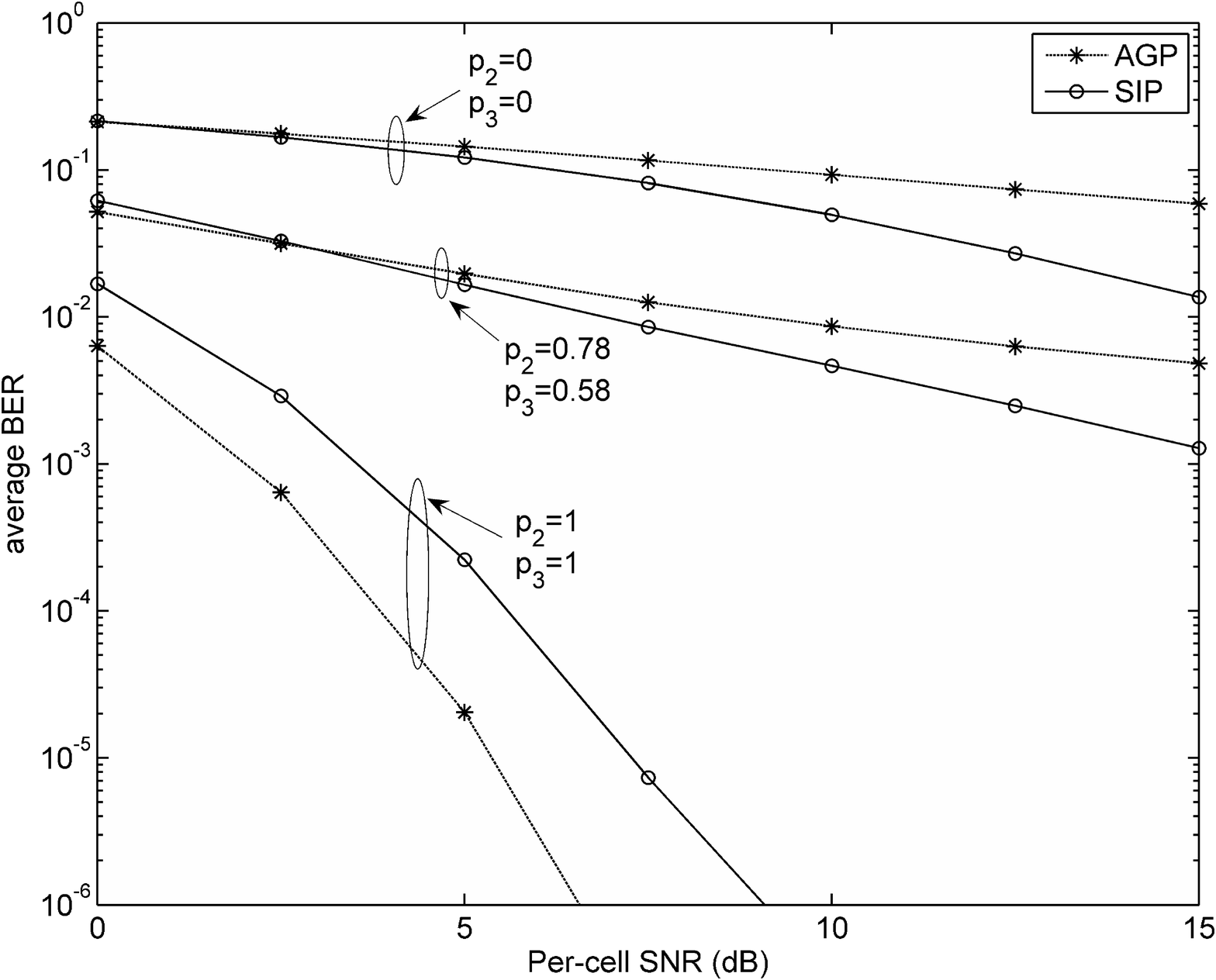} 
\vspace{-0.5em}
\renewcommand{\figurename}{Fig.}\caption{Average BER for $B=3$, $N_{\mathrm{R}}=4$ (perfect feedback).}
\end{figure}
\begin{figure}[H]
\centering\includegraphics[width=8cm]{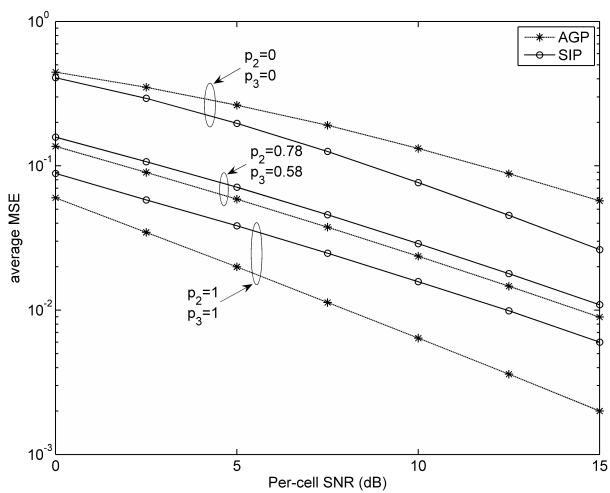} 
\vspace{-0.5em}
\renewcommand{\figurename}{Fig.}\caption{Average MSE for $B=3$, $N_{\mathrm{R}}=2$ (perfect feedback).}
\end{figure}

\subsection{Peformance of Finite Rate Feedback Systems}

Since the proposed SIP scheme is developed under the assumption of
perfect feedback, it is highly motivated to investigate whether the
SIP scheme still works in case of practical finite-rate feedback systems,
i.e., the indices of the optimal codeword $\mathbf{W}_{b}^{\textrm{opt,cb}}$s
rather than $\mathbf{W}_{b}^{\textrm{opt}}$s themselves are fed back.
First, we investigate the choice of $d$ in the codebook based feedback.
In Fig. 15 and 16, average BER performance of the SIP and AGP schemes
are shown for $B=2$ $\left(p_{2}=0.78\right)$ or $B=3$ $\left((p_{2},p_{3})=(0.78,\,0.58)\right)$,
$N_{\textrm{R}}=2$ and $d=1\sim5$. We can observe in both figures
that the performance gain suffers from a diminishing return as $d$
increases and $d=4$ seems to be a good tradeoff between performance
improvement and feedback overhead. Moreover, $d=4$ is a common assumption
for codebook designs for BSs with 4 transmit antennas in the LTE-A
system \cite{LTE-A[3]}. Thus in the following we provide new simulation
results plotted in Fig. 17 to 24 for $d=4$ to illustrate the performance
degradation because of limited-bit feedback in compare with Fig. 6
to 13. In our simulations, precoder candidates in the codebook are
randomly generated as matrix composed of orthogonal normalized vectors
\cite{RVQ[35]} for each channel realization. It can be observed from
these figures that although the performance degradation is notable
the gains of the SIP scheme shown in Fig. 6 to 13 are safely preserved.

\begin{figure}[H]
\centering\includegraphics[width=8cm]{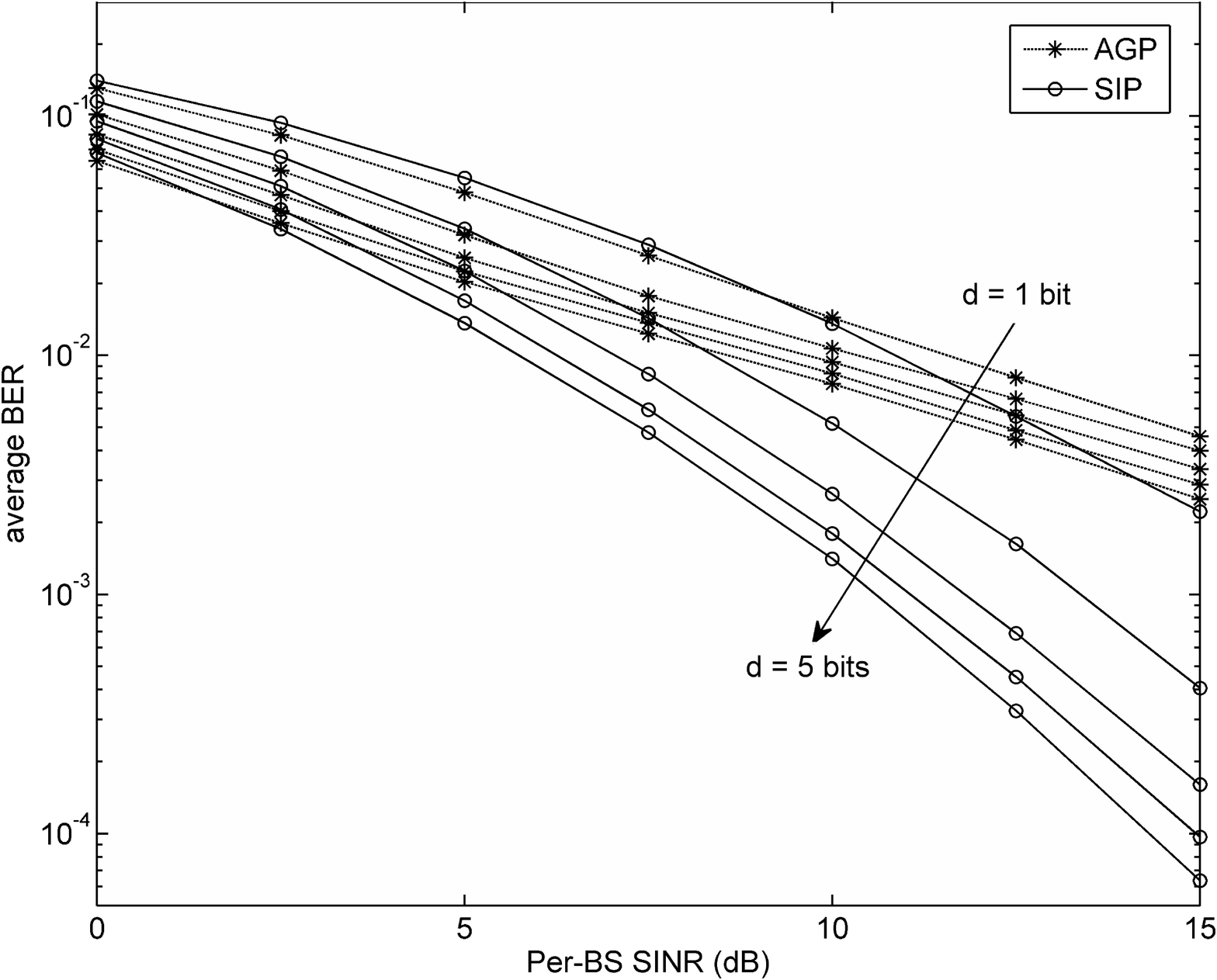} 
\vspace{-0.5em}
\renewcommand{\figurename}{Fig.}\caption{Average BER for $B=2$, $N_{\mathrm{R}}=2$, $d=1\sim5$ (codebook
based feedback).}
\end{figure}
\begin{figure}[H]
\centering\includegraphics[width=8cm]{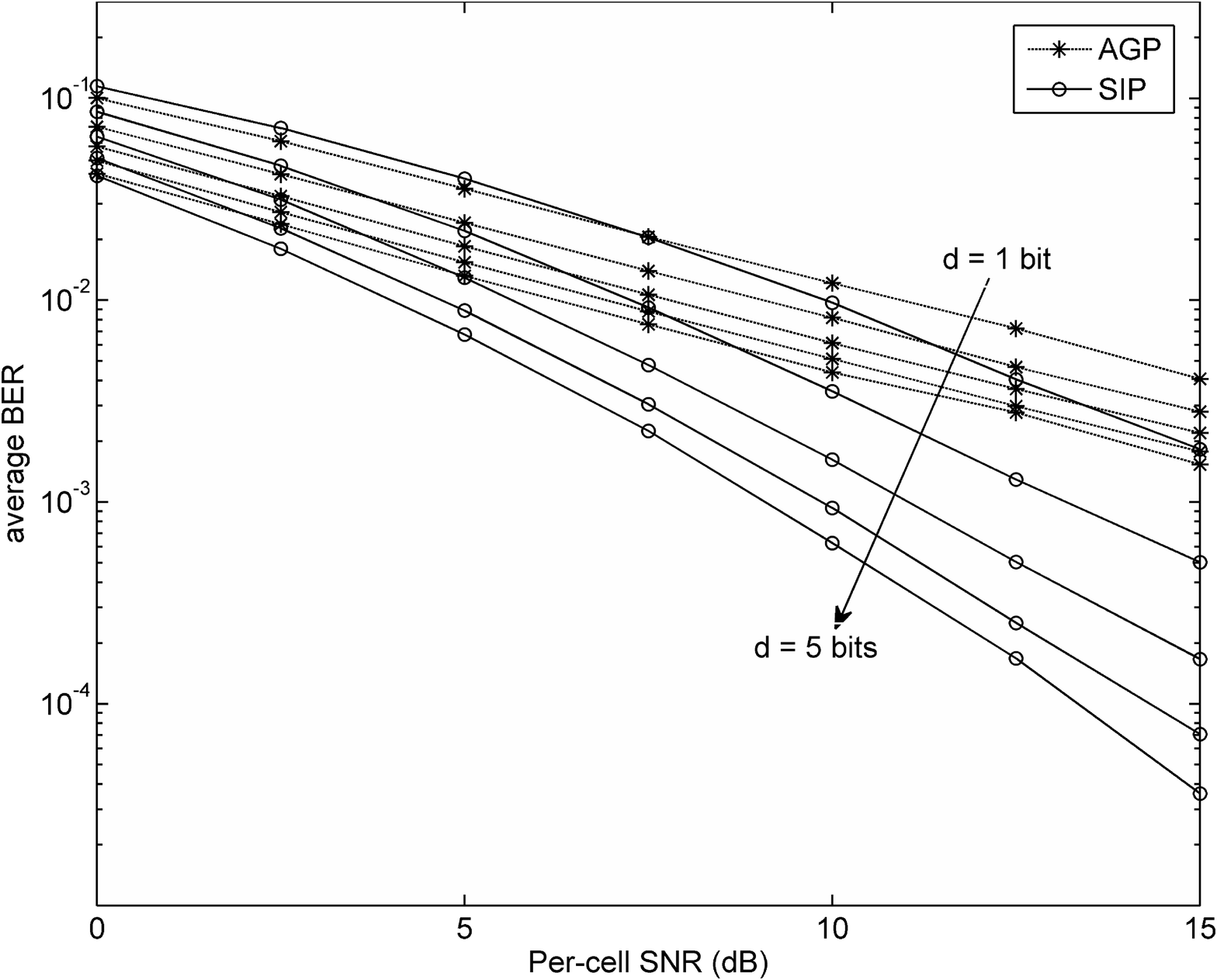} 
\vspace{-0.5em}
\renewcommand{\figurename}{Fig.}\caption{Average BER for $B=3$, $N_{\mathrm{R}}=2$, $d=1\sim5$ (codebook
based feedback).}
\end{figure}
\begin{figure}[H]
\centering\includegraphics[width=8cm]{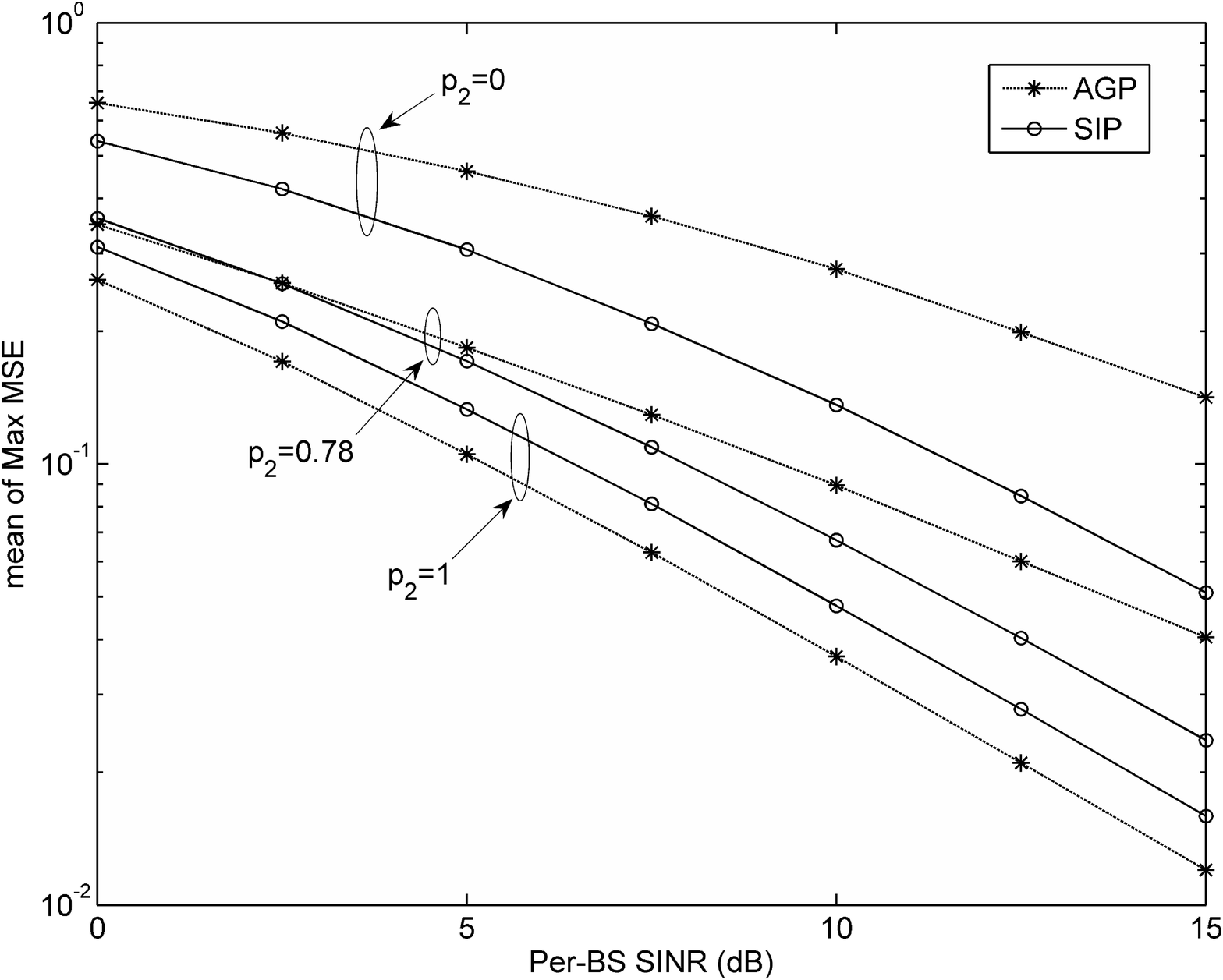} 
\vspace{-0.5em}
\renewcommand{\figurename}{Fig.}\caption{Mean of the maximum of sub-stream MSEs for $B=2$, $N_{\mathrm{R}}=2$,
$d=4$ (codebook based feedback).}
\end{figure}
\begin{figure}[H]
\centering\includegraphics[width=8cm]{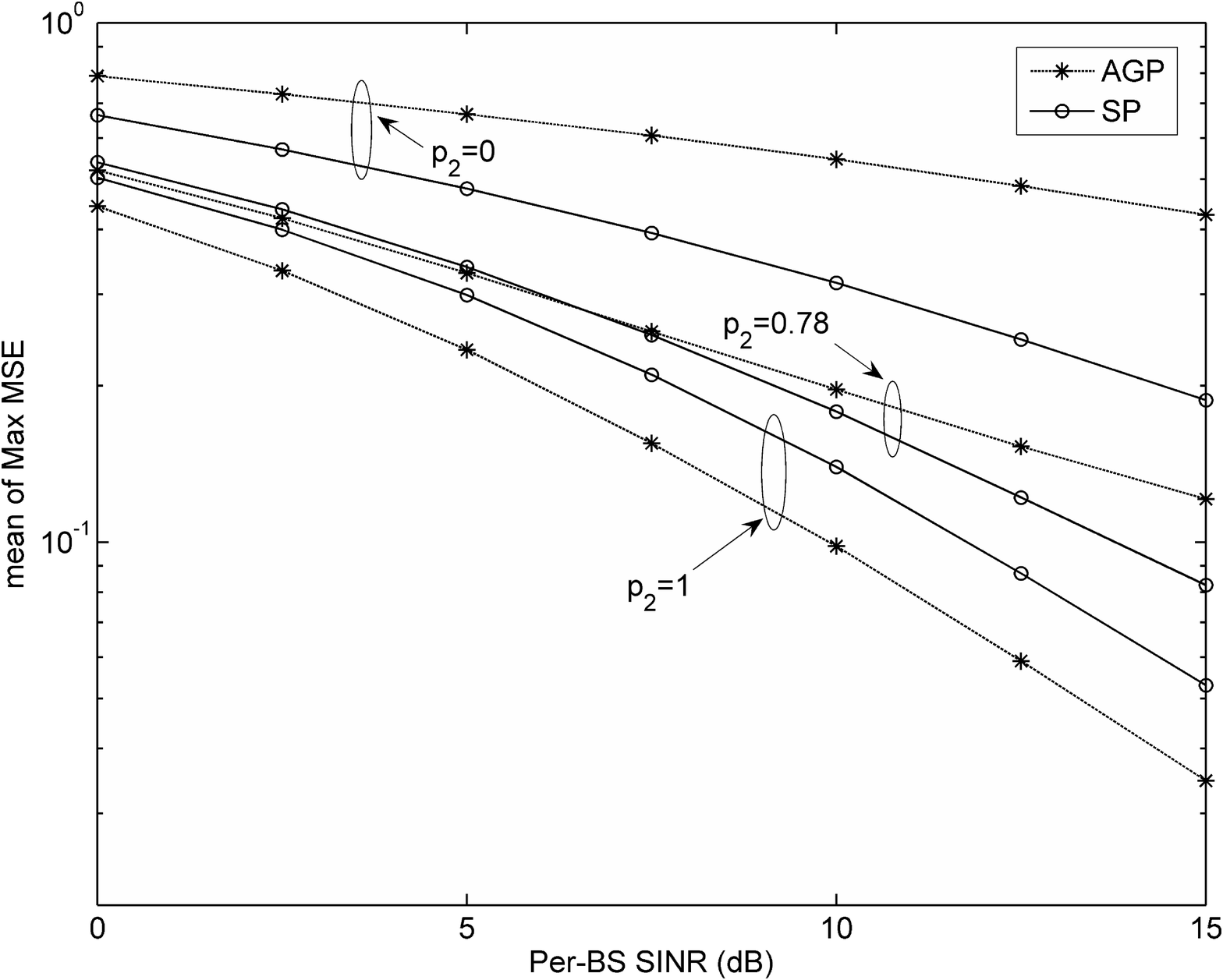} 
\vspace{-0.5em}
\renewcommand{\figurename}{Fig.}\caption{Mean of the maximum of sub-stream MSEs for $B=2$, $N_{\mathrm{R}}=4$,
$d=4$ (codebook based feedback).}
\end{figure}
\begin{figure}[H]
\centering\includegraphics[width=8cm]{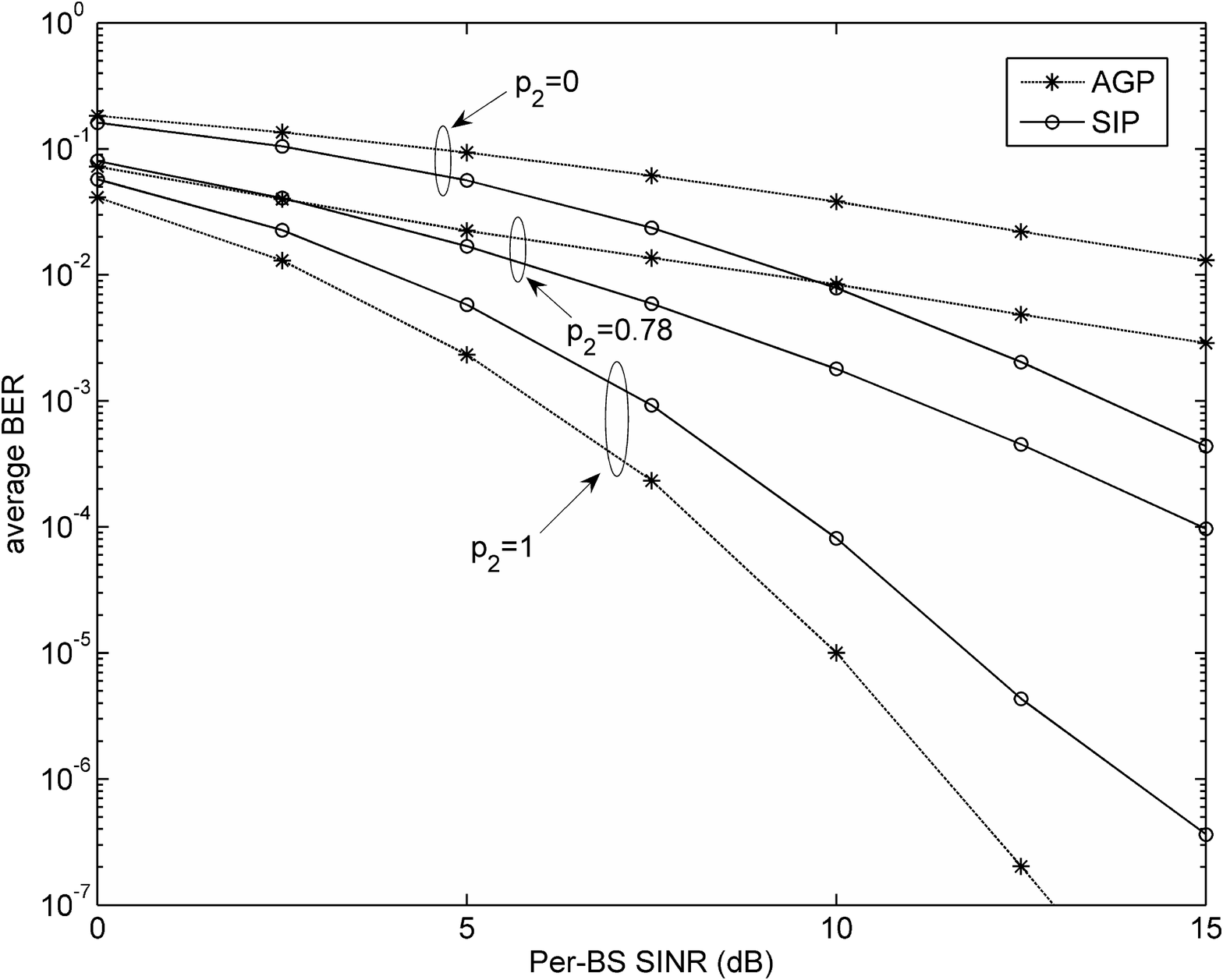} 
\vspace{-0.5em}
\renewcommand{\figurename}{Fig.}\caption{Average BER for $B=2$, $N_{\mathrm{R}}=2$, $d=4$ (codebook based
feedback).}
\end{figure}
\begin{figure}[H]
\centering\includegraphics[width=8cm]{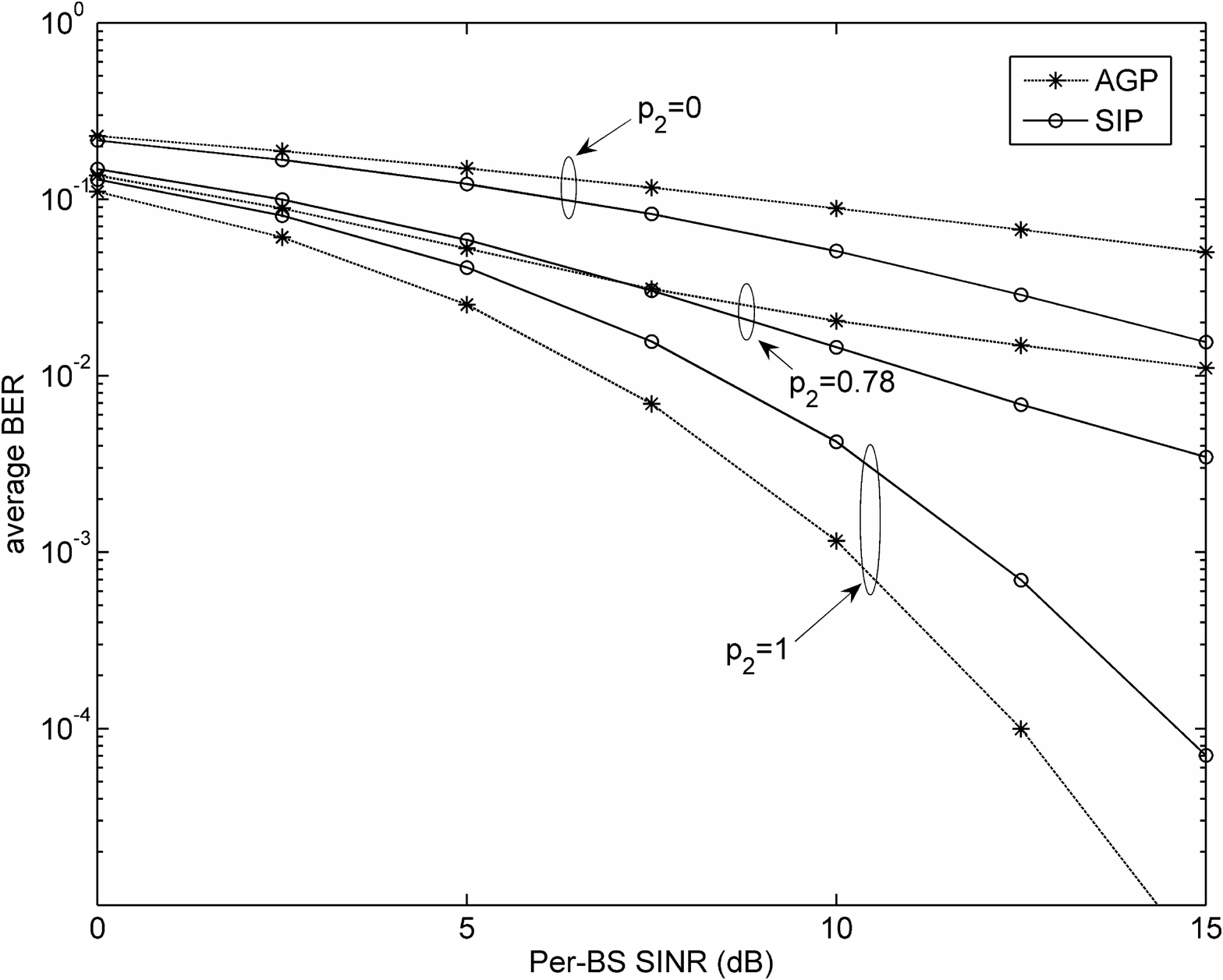} 
\vspace{-0.5em}
\renewcommand{\figurename}{Fig.}\caption{Average BER for $B=2$, $N_{\mathrm{R}}=4$, $d=4$ (codebook based
feedback).}
\end{figure}
\begin{figure}[H]
\centering\includegraphics[width=8cm]{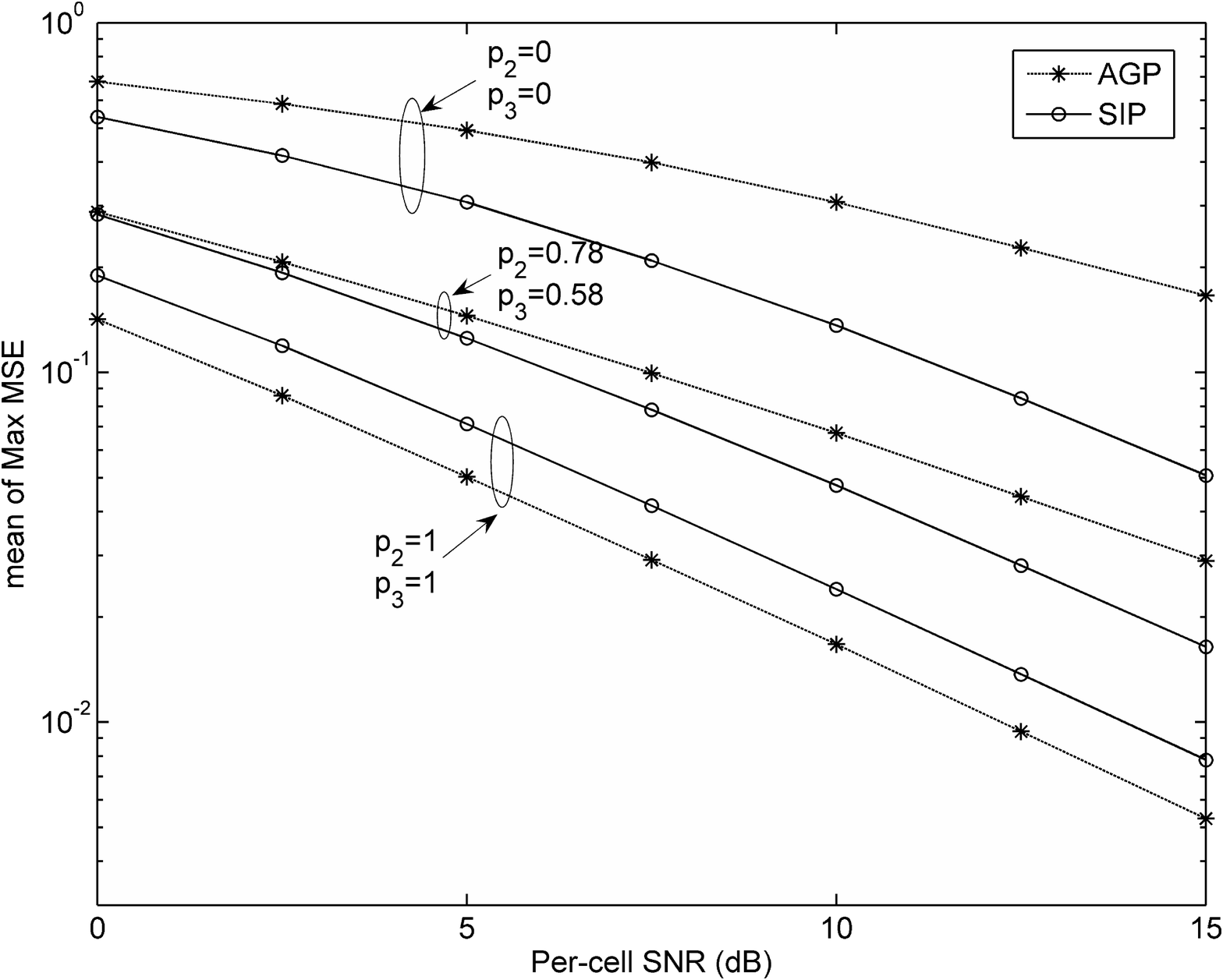} 
\vspace{-0.5em}
\renewcommand{\figurename}{Fig.}\caption{Mean of the maximum of sub-stream MSEs for $B=3$, $N_{\mathrm{R}}=2$,
$d=4$ (codebook based feedback).}
\end{figure}
\begin{figure}[H]
\centering\includegraphics[width=8cm]{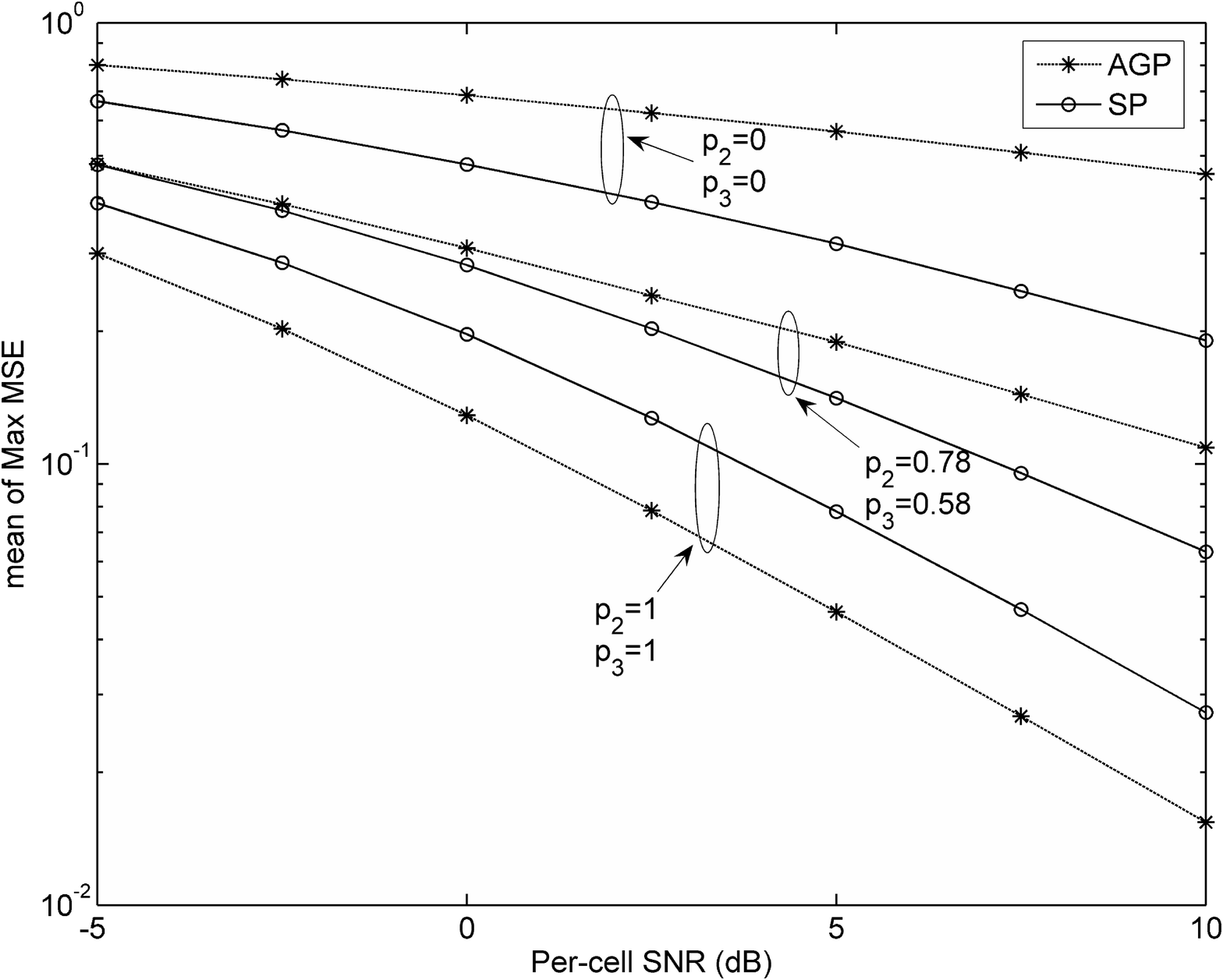} 
\vspace{-0.5em}
\renewcommand{\figurename}{Fig.}\caption{Mean of the maximum of sub-stream MSEs for $B=3$, $N_{\mathrm{R}}=4$,
$d=4$ (codebook based feedback).}
\end{figure}
\begin{figure}[H]
\centering\includegraphics[width=8cm]{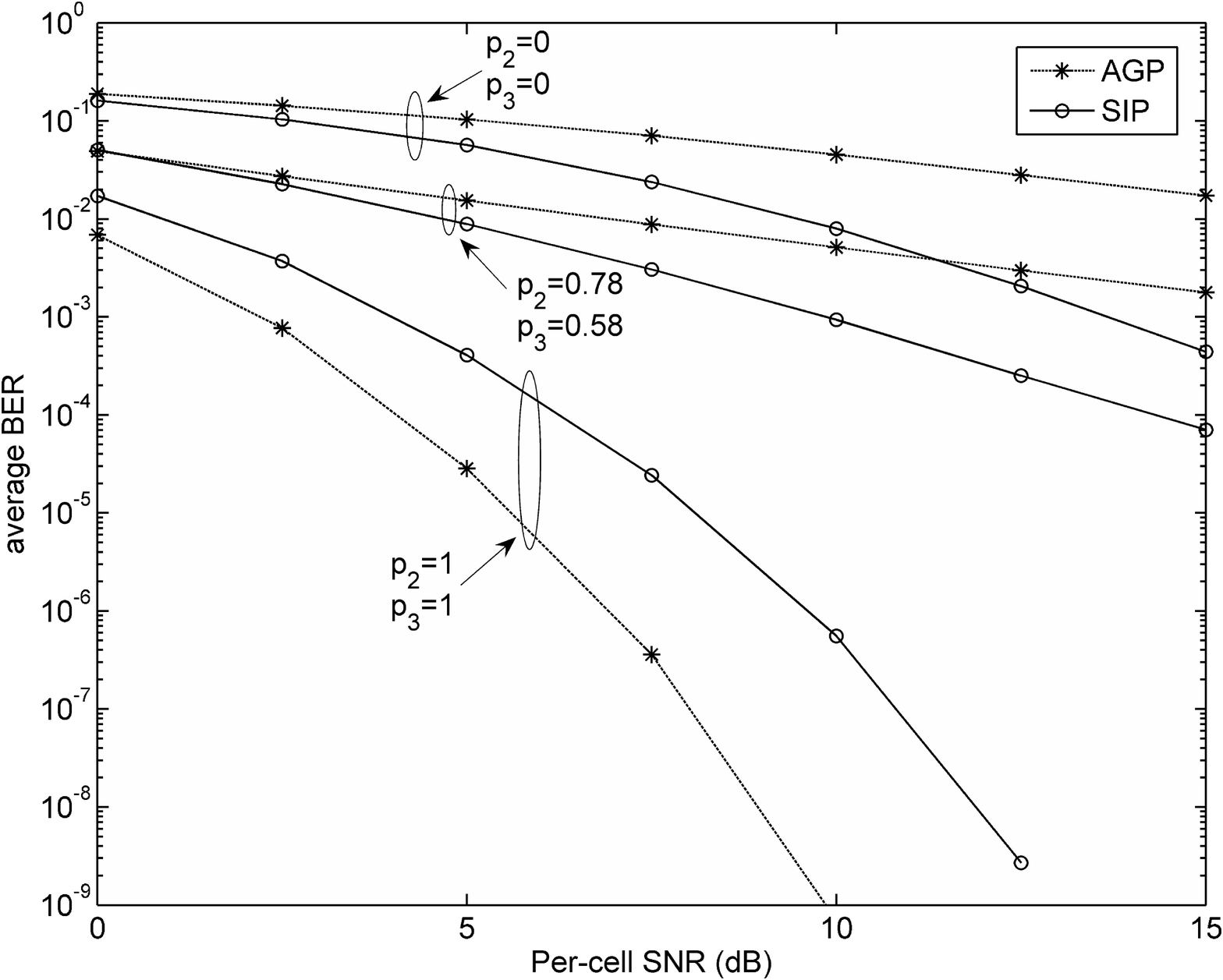} 
\vspace{-0.5em}
\renewcommand{\figurename}{Fig.}\caption{Average BER for $B=3$, $N_{\mathrm{R}}=2$, $d=4$ (codebook based
feedback).}
\end{figure}
\begin{figure}[H]
\centering\includegraphics[width=8cm]{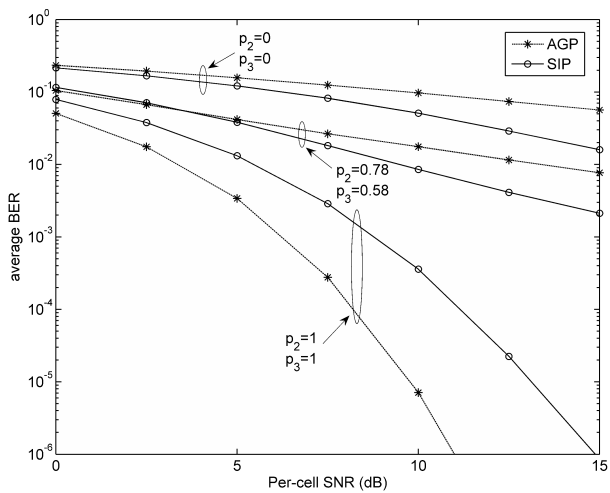} 
\vspace{-0.5em}
\renewcommand{\figurename}{Fig.}\caption{Average BER for $B=3$, $N_{\mathrm{R}}=4$, $d=4$ (codebook based
feedback).}
\end{figure}

\subsection{Extension to the Multi-UE Scenario}

Finally, we briefly discuss the further extension of the proposed
SIP scheme to the multi-UE scenarios and more sophisticated evaluations
using system-level simulations. When multiple UEs are involved, the
Wiener filter in (18) can be replaced by a block diagonal matrix to
reflect that individual UE\textquoteright{}s receive processing capability
is captured by the corresponding block matrix and no receive cooperation
is assumed among different UEs. Then the precoders form multiple UEs
can be derived using the proposed SIP scheme. However, the closed-form
expression for $\mathbf{W}_{1}^{\textrm{opt}}$ shown by (15) no longer
exists. The algorithms for optimizing $\mathbf{W}_{1}^{\textrm{opt}}$
with various objective functions can be found in \cite{motivation_MINMAX[15]},
which is beyond the scope of this paper and will be considered in
the future works. Besides, the simulation results given in this paper
are on a link-level basis. More sophisticated evaluations using system-level
simulations \cite{DPS_muting} considering cell-edge/cell-average
spectral efficiency, UE distribution, multi-cell scheduler, retransmission
in case of packet error, traffic modeling, propagation channel modeling,
channel estimation errors, BS power settings, etc., are beneficial
to the investigation of the performance gain offered by the proposed
scheme in more practical scenarios. Thus we will improve our simulation
methods in the future works.

\section{Conclusion}

In this paper, transmission precoder design based on a sequential
and incremental approach for JT network MIMO systems with imperfect
backhaul is studied. The conventional autonomous global precoding
(AGP) scheme suffers from severe performance degradation in the event
of partial JT and ST resulted from imperfect backhaul communications.
A sequential and incremental precoding (SIP) scheme is proposed to
overcome the drawbacks of the existing schemes. The key problem is
first illustrated and solved with a two-BS JT system, and the results
are then generalized to multi-BS JT network MIMO systems. Simulation
results show that our scheme significantly outperforms the AGP scheme
when practical backhaul link is considered. Finally, future works
of extending the proposed SIP scheme to multi-UE scenarios and more
sophisticated evaluations using system-level simulations are briefly
discussed.

\end{document}